\documentclass[a4paper,11pt]{article}
\pdfoutput=1 
\usepackage{verbatim}

\usepackage{jheppub}
\usepackage{amsmath,amsfonts,amssymb,amsthm,color}
\usepackage[T1]{fontenc} 
\usepackage{enumerate}
\usepackage{tikz, tikz-3dplot, pgfplots}
\usepackage[utf8]{inputenc}
\usepackage{slashed}
\usepackage{booktabs}
\usepackage{caption}
\usepackage{subcaption}
\usepackage{footmisc}
\usepackage{multirow,graphicx}

\usepackage{float}
\usepackage{changepage}
\usepackage{relsize}
\usepackage{enumitem}

\usepackage{array}

\usepackage{changepage}
\usepackage{float}
\usepackage{tikz}
\usetikzlibrary{decorations.text}
\usepackage{enumitem}
\setlist{itemsep=0pt}
\usetikzlibrary{math}

\usepackage{tikz}
\usepackage{pgfplots}
\usetikzlibrary{decorations.text,intersections, pgfplots.fillbetween}
\usetikzlibrary{math}
 \usetikzlibrary{snakes,3d,shapes.geometric,shadows.blur}
\usepackage{tikz-3dplot}

\definecolor{amaranthred}{rgb}{0.83,0.13,0.18}
\definecolor{amazon}{rgb}{0.23,0.48,0.34}
\definecolor{bdazzledblue}{rgb}{0.18,0.35,0.58}
\definecolor{absolutezero}{rgb}{0.0,0.28,0.73}
\definecolor{bitterlemon}{rgb}{0.79,0.88,0.05}
\definecolor{byzantine}{rgb}{0.74,0.2,0.64}
\definecolor{turquoise}{rgb}{0.19, 0.84, 0.78}

\newcolumntype{C}{>{\centering\arraybackslash}m{20em}}

\newcolumntype{M}[1]{>{\centering\arraybackslash}m{#1}}

\newcommand{\comm}[1]{} 

\def\({\left(}
\def\){\right)}
\def\[{\left[}
\def\]{\right]}

\def\One{{\hbox{ 1\kern-.8mm l}}}

\def\barray{\begin{array}}
\def\earray{\end{array}}
\def\be{\begin{equation}}
\def\ee{\end{equation}}
\def\bea{\begin{eqnarray}}
\def\eea{\end{eqnarray}}
\def\bal{\begin{align}}
\def\eal{\end{align}}
\def\nn{\nonumber}

\def\R{R_y} 

\def\-{\,-\,}
\def\={\,=\,}
\def\+{\,+\,}
\def\equi{\,\equiv\,}

\numberwithin{equation}{section} 

\definecolor{cardinal}{rgb}{0.6,0,0}
\definecolor{darkgreen}{rgb}{0,0.4,0}
\definecolor{golden}{rgb}{0.92, 0.7, 0}
\definecolor{midnight}{rgb}{0, 0, 0.5}
\definecolor{darkblue}{rgb}{0, 0, 0.7}
\definecolor{purple}{rgb}{0.5, 0, 0.5}

\definecolor{amaranthred}{rgb}{0.83,0.13,0.18}
\definecolor{amazon}{rgb}{0.23,0.48,0.34}
\definecolor{bdazzledblue}{rgb}{0.18,0.35,0.58}
\definecolor{absolutezero}{rgb}{0.0,0.28,0.73}
\definecolor{bitterlemon}{rgb}{0.79,0.88,0.05}
\definecolor{byzantine}{rgb}{0.74,0.2,0.64}
\definecolor{turquoise}{rgb}{0.19, 0.84, 0.78}
\definecolor{burgundy}{rgb}{0.5, 0.0, 0.13}

\def\IR{\mathbb{R}}

\def\cI{{\cal I}}
\def\cJ{{\cal J}}

\def\cM{{\cal M}}
\def\cN{{\cal N}}
\def\cP{{\cal P}}
\def\cO{{\cal O}}

\def\cS{{\cal S}}

\def\cW{{\cal W}}
\def\cX{{\cal X}}

\def\cZ{{\cal Z}}

\makeatletter
\newcommand\footnoteref[1]{\protected@xdef\@thefnmark{\ref{#1}}\@footnotemark}
\makeatother

\usetikzlibrary{positioning,calc,fadings,decorations.pathreplacing}

\tikzset{
 diffuse color/.initial = black,                       
}

\tikzset{
 linear opacity/.initial=0.5,                          
 linear stroke/.style = {                              
   preaction={                                         
     draw=\pgfkeysvalueof{/tikz/diffuse color},        
     line width = (2.0-#1)*\pgflinewidth,              
     opacity=\pgfkeysvalueof{/tikz/linear opacity},white}},  
 diffuse gradient/.style={                             
   draw = none,                                        
   linear opacity=#1,                                  
   linear stroke/.list={0.0,#1,...,1.0}},              
 diffuse gradient/.default=1,                          
}

\tikzset{
 non-linear stroke/.style = {                          
   preaction={                                         
     draw=\pgfkeysvalueof{/tikz/diffuse color},        
     line width = (2.0-#1)*\pgflinewidth,              
     opacity=#1,white}},                                     
 diffuse falloff/.style={                              
   draw = none,                                        
   non-linear stroke/.list={0.0,#1,...,1.0}},          
 diffuse falloff/.default=1,                           
}
 
\newcommand\pgfmathsinandcos[3]{%
  \pgfmathsetmacro#1{sin(#3)}%
  \pgfmathsetmacro#2{cos(#3)}%
}
\newcommand\LongitudePlane[3][current plane]{%
  \pgfmathsinandcos\sinEl\cosEl{#2} 
  \pgfmathsinandcos\sint\cost{#3} 
  \tikzset{#1/.style={cm={\cost,\sint*\sinEl,0,\cosEl,(0,0)}}}
}
\newcommand\LatitudePlane[3][current plane]{%
  \pgfmathsinandcos\sinEl\cosEl{#2} 
  \pgfmathsinandcos\sint\cost{#3} 
  \pgfmathsetmacro\yshift{\cosEl*\sint}
  \tikzset{#1/.style={cm={\cost,0,0,\cost*\sinEl,(0,\yshift)}}} %
}

\newcommand\DrawLatitudeCircle[2][1]{
  \LatitudePlane{\angEl}{#2}
  \tikzset{current plane/.prefix style={scale=#1}}
  \pgfmathsetmacro\sinVis{sin(#2)/cos(#2)*sin(\angEl)/cos(\angEl)}
  \pgfmathsetmacro\angVis{asin(min(1,max(\sinVis,-1)))}
  \draw[current plane] (\angVis:1) arc (\angVis:-\angVis-180:1);
  \draw[current plane,dashed] (180-\angVis:1) arc (180-\angVis:\angVis:1);
}

\tikzset{%
  >=latex, 
  inner sep=0pt,%
  outer sep=2pt,%
  mark coordinate/.style={inner sep=0pt,outer sep=0pt,minimum size=3pt,
    fill=black,circle}%
}

 \newcommand{\bubble}{
    \begin{tikzpicture}[remember picture,trim left=0.1cm]

\def\R{1.1} 
\def\angEl{25} 
\def\angAz{-105} 
\def\angPhi{-40} 
\def\angBeta{19} 

\pgfmathsetmacro\H{\R*cos(\angEl)} 
\tikzset{xyplane/.style={cm={cos(\angAz),sin(\angAz)*sin(\angEl),-sin(\angAz),
                              cos(\angAz)*sin(\angEl),(0,-\H)}}}
\LongitudePlane[xzplane]{\angEl}{\angAz}
\LongitudePlane[pzplane]{\angEl}{\angPhi}
\LatitudePlane[equator]{\angEl}{0}

\fill[ball color=white] (0,0) circle (\R); 
\draw (0,0) circle (\R);

\draw[->,white!70!black] (0,-1.3*\H) -- (0,1.5*\R) node[above] {};

\coordinate[gray] (O) at (0,0);
\coordinate[mark coordinate] (N) at (0,\H);
\coordinate[mark coordinate] (S) at (0,-\H);
\coordinate (P) at (\H,0);
\path[pzplane] (\R,0) coordinate (PE);
\path[xzplane] (\R,0) coordinate (XE);

\DrawLatitudeCircle[\R]{0} 

\draw[->,line width=2pt, -latex, red!70!black] (-1,\H) to [bend right=40] (1.2,\H+0.07);
\draw[red!70!black] (1.2,\H+0.07) node[above=0.0] {};

\draw (0,-\H-0.9) node[above=0.0] {\small $k\in \mathbb{N}$; $(\ell,N)\in \mathbb{Z}$; $|q|<1$};

\draw[<-] (0+0.1,\H+0.01) -- (1.7,\H+0.25) node[above,right] {\scriptsize $\IR^4/\mathbb{Z}_{|\ell-N k|}$};
\draw[<-] (0+0.1,-\H-0.01) -- (1.7,-\H-0.25) node[above,right] {\scriptsize $\IR^4/\mathbb{Z}_{|\ell|}$};

\draw[thin,decorate,decoration={brace,raise=0.5pt,amplitude=1ex}] (1.3*\R,\H) -- (1.3*\R,-\H)
    node[midway, right=0.3] {\footnotesize S$^2$ bubble at an origin of $\IR^2$}; 

 \end{tikzpicture}}

\begin{document}

\phantom{AAA}
\vspace{-10mm}

\begin{flushright}
%
%
\end{flushright}

\vspace{2cm}

\begin{center}

{\fontsize{19}{23}\selectfont{\bf Asymptotically Flat Rotating Topological Stars}}

\vspace{0.2cm}

\vspace{1cm}

{\bf \normalsize Pierre Heidmann,$^{1}$ Paolo Pani,$^{2}$ and Jorge E. Santos$^{3}$}
\vspace{0.5cm}\\

\centerline{$^1$ Department of Physics and Center for Cosmology and AstroParticle Physics (CCAPP),}
\centerline{The Ohio State University, Columbus, OH 43210, USA}
\centerline{$^2$ Dipartimento di Fisica, Sapienza Università di Roma \& INFN, Sezione di Roma,}
\centerline{Piazzale Aldo Moro 5, 00185, Roma, Italy}
\centerline{$^3$ Department of Applied Mathematics and Theoretical Physics,}
\centerline{University of Cambridge, Wilberforce Road, Cambridge, CB3 0WA, UK}

\vspace{0.5 cm}

{\footnotesize\upshape\ttfamily heidmann.5@osu.edu, paolo.pani@unirom1.it, jss55@cam.ac.uk} 

 \vspace{0.5 cm}

\end{center}

\begin{adjustwidth}{5mm}{5mm} 
\begin{abstract}
\noindent
We construct a new class of smooth, horizonless, non-supersymmetric solutions in five-dimensional minimal supergravity, which we call \emph{rotating topological stars}.
Built from a Kerr-Taub-bolt geometry embedded in five dimensions, they constitute the first rotating generalization of the topological star compatible with both smoothness in the interior and standard Kaluza-Klein asymptotics, S$^1\times\mathbb{R}^{1,3}$.
The solutions carry angular momentum, magnetic and electric charges, and form a discrete tower of states labeled by a primary quantum number controlling the spin.
Remarkably, despite lying outside the black-hole extremality bound, they can approach arbitrarily closely (in conserved charges) the Kerr black string with a large boost along the fifth dimension, making them relevant prototypes for rotating and astrophysical black-hole microstates.
We analyze their geometry in detail, including  their gravitational multipoles that can significantly deviate from those of black holes and the presence of an ergoregion, and show that both geodesics and scalar perturbations separate, paving the way for analyzing their dynamics in future work.

\end{abstract}
\end{adjustwidth}

\vspace{8mm}
 

\thispagestyle{empty}

\newpage



\tableofcontents

\flushbottom

\begin{adjustwidth}{-0.8cm}{-0.8cm}

\section{Introduction}

Understanding the fundamental structure of black holes remains one of the most pressing challenges in quantum gravity. Arguments from quantum information theory~\cite{Mathur:2009hf,Almheiri:2012rt} and from explicit constructions in string theory~\cite{Mathur:2005zp,Bena:2022rna,Bena:2022ldq} both indicate that the enormous microstructure accounting for the Bekenstein-Hawking entropy must be encoded in the vicinity of the horizon. This suggests that new physics, beyond the reach of General Relativity, may emerge at the horizon scale with potentially observable signatures~\cite{Mayerson:2020tpn}.

While generic black-hole microstates are intrinsically quantum, some can be sufficiently coherent to admit a classical description as gravitational solitons. Such solutions are as compact as black holes but replace the horizon with smooth, horizonless structures, thereby capturing the large-scale features of black-hole microstructure at the horizon scale. In string theory, many coherent microstates have been constructed for supersymmetric black holes~\cite{Bena:2022rna,Bena:2025pcy,Heidmann:2019xrd,Shigemori:2019orj}, theorizing the only viable gravitational mechanism for sustaining smooth horizon-scale structure with a vast phase space: nontrivial topology induced by the deformation of extra compact dimensions and supported by electromagnetic flux~\cite{Gibbons:2013tqa}.

A crucial ingredient in these constructions is the role of compact dimensions and nontrivial topological structures in spacetime, which enables smooth caps that resolve the curvature singularities of black holes at their horizon. Although microstate geometries are typically formulated within ten- or eleven-dimensional supergravity, consistent dimensional reductions often retain the essential physical features --~such as smoothness, conserved charges, and nontrivial topology. In particular, five-dimensional supergravity (from M-theory on T$^6$) has proved to be a powerful setting for building regular, asymptotically S$^1\times\mathbb{R}^{1,3}$ solitons~\cite{Bena:2007kg,Heidmann:2021cms,Chakraborty:2025ger}.

While supersymmetric microstate geometries have been central in probing black-hole physics beyond General Relativity~\cite{Tyukov:2017uig,Bena:2020iyw,Bena:2019azk,Martinec:2020cml,Bena:2020yii,Chakrabarty:2021sff,Bianchi:2022qph}, supersymmetric models alone cannot describe realistic astrophysical black holes. A \emph{key goal} in this program is to construct and analyze coherent microstate geometries of astrophysically relevant, nonextremal black holes. Such black holes are typically (1) rotating and (2) neutral, lying as far from the BPS bound as possible. Achieving this, however, requires tackling the full nonlinear structure of Einstein’s equations within supergravity.

Recent progress shows that this objective is within reach, successfully addressing point (2). Integrable approaches to supergravity~\cite{Bah:2021owp,Heidmann:2021cms,Chakraborty:2025ger} have enabled the explicit construction of nonextremal static topological solitons, including coherent microstates of the Schwarzschild black hole~\cite{Bah:2022yji,Bah:2023ows} and analytically tractable charged configurations such as the topological star~\cite{Bah:2020ogh,Bah:2020pdz} and the $\cW$-soliton~\cite{Chakraborty:2025ger,Dima:2025tjz}.
These solutions demonstrate how flux and nontrivial topology can sustain horizon-scale structure far beyond the BPS limit, but they also reveal a key challenge: all currently known examples are static, and simple attempts to add rotation have so far failed.

A number of rotating, nonextremal topological solitons in supergravity are known, most notably the JMaRT geometry~\cite{Jejjala:2005yu} — asymptotic to $\IR^{1,4}\times$S$^1$ rather than $\IR^{1,3}\times$S$^1$ — together with several multicenter generalizations~\cite{Bossard:2014yta,Bossard:2014ola,Bena:2015drs,Bena:2016dbw,Bossard:2017vii} and a four-dimensional analogue~\cite{Giusto:2007tt}.
However, these solutions rely on angular momentum well above the black-hole bound to counterbalance the absence of electromagnetic repulsion for non-BPS configurations.
Recently, a rotating generalization of the topological star was obtained in~\cite{Bianchi:2025uis}, but it fails to admit standard Kaluza-Klein asymptotics, instead approaching $(\IR^{1,3}\times$S$^1)/\mathbb{Z}_q$, corresponding to a S$^1$ over a magnetic Melvin universe \cite{Dowker:1995gb}.

\subsection{Summary of the results}

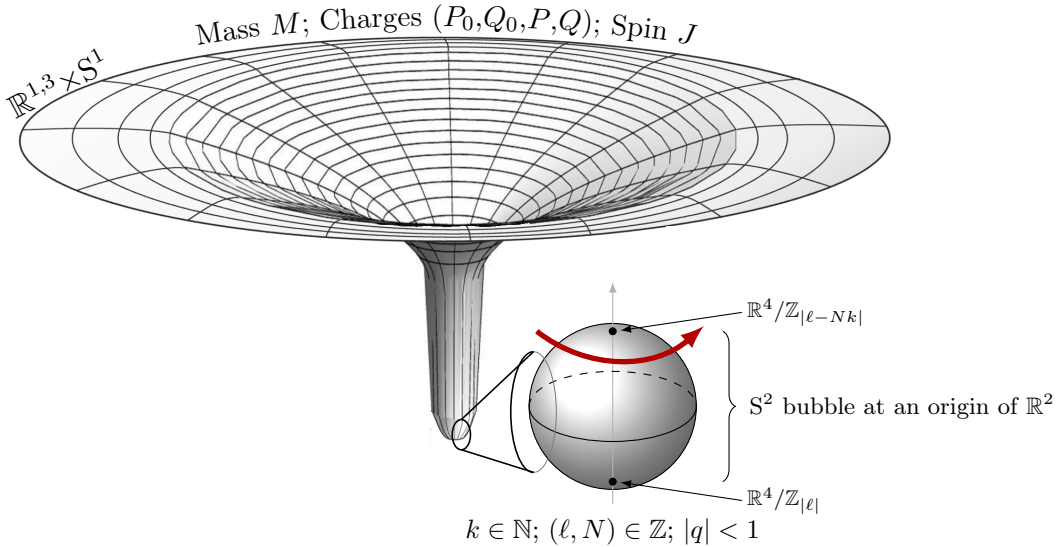
\begin{figure}[t]
\begin{center}
\begin{tikzpicture}
\tikzmath{\xxpic = -4; \ycap=-2.3;\sizefirstlens=0.2;\xleftfirstlens=0.05;\xrightfirstlens=1.;\yrightfirstlens=0.3;\sizerightfirstlens=0.8;\xbubble=5;\ybubble=0.3;\sizeseclens=0.23;\xleftseclens=\xbubble-0.6;\yrightseclens=1.15;\sizerightseclens=1;\xrightseclens=\xbubble+1.3;} 

\node[inner sep=0pt] (bubblefar) at (\xxpic,0)
    {\includegraphics[width=0.8\textwidth]{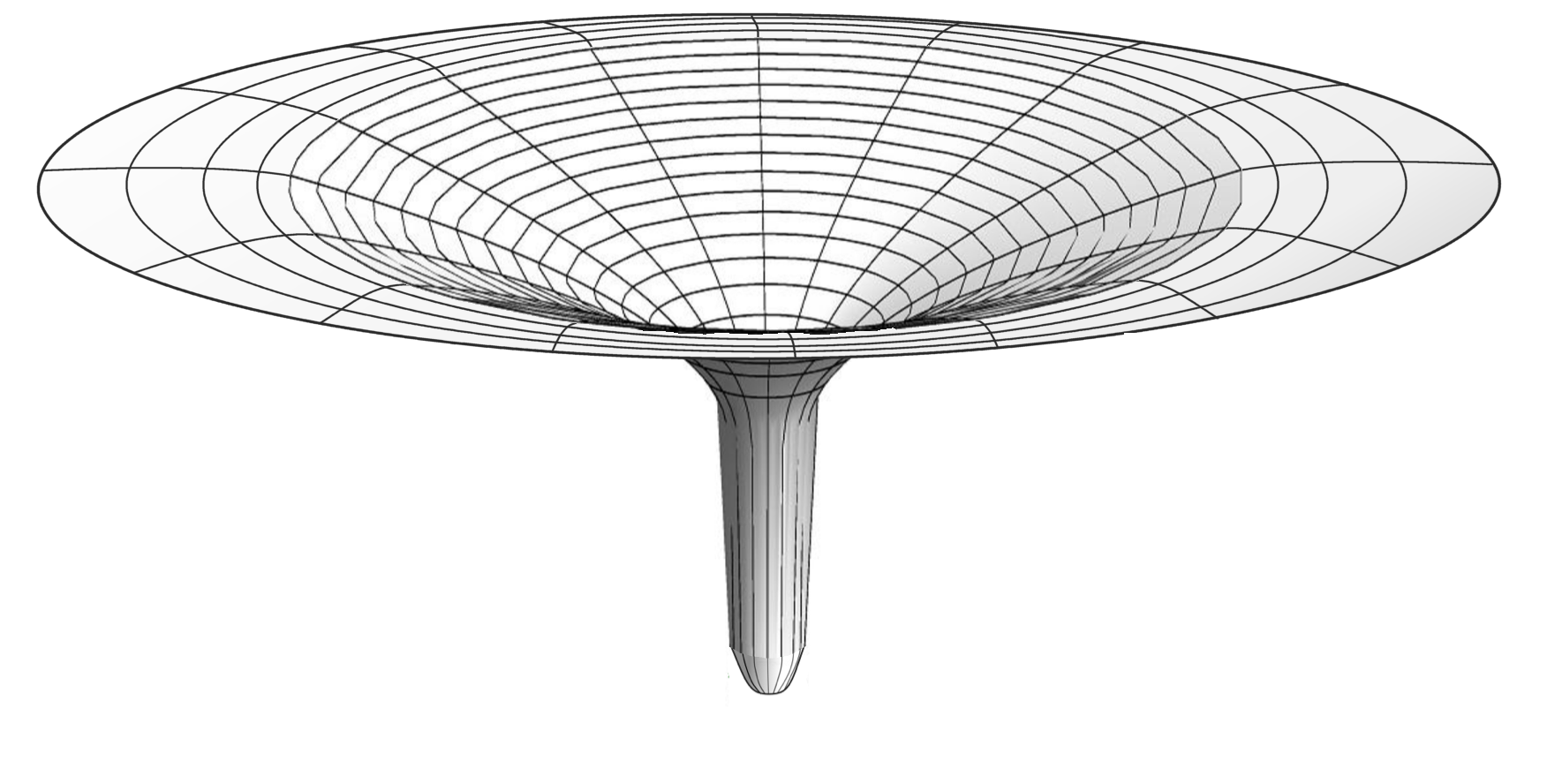}};
\node[rotate=28] at (\xxpic-5.25,2.35)  {\large $\mathbb{R}^{1,3}\times$S$^1$} ;
    
\node (bubble) at (\xxpic+\xbubble,\ycap+\ybubble)
    {\bubble};
    

\draw[decorate,decoration={text along path,
    text={Mass $M$; Charges $({P_0},{Q_0},P,Q)$; Spin $J$},
    text align=center,
    raise=2pt 
}] 
(\xxpic-6.05,\ycap+3.85) arc[start angle=180, end angle=0, x radius=5.9, y radius=1.5];

	\draw[color=gray] (\xxpic+\xrightfirstlens,\ycap+\yrightfirstlens-\sizerightfirstlens) arc (-90:90:0.3 and \sizerightfirstlens);
	\draw[semithick] (\xxpic+\xrightfirstlens,\ycap+\yrightfirstlens-\sizerightfirstlens) -- (\xxpic+\xleftfirstlens,\ycap-\sizefirstlens);
	\draw[semithick] (\xxpic+\xrightfirstlens,\ycap+\yrightfirstlens+\sizerightfirstlens) -- (\xxpic+\xleftfirstlens,\ycap+\sizefirstlens);
	\draw[semithick] (\xxpic+\xrightfirstlens,\ycap+\yrightfirstlens-\sizerightfirstlens) arc (270:90:0.3 and \sizerightfirstlens);
	\draw[semithick] (\xxpic+\xleftfirstlens,\ycap) ellipse (0.12 and 0.2);

\end{tikzpicture}
\caption{Schematic description of a rotating topological star in $\IR^{1,3}\times$S$^1$.}
\label{fig:Schematic}
\end{center}
\end{figure}

We take a substantial step toward constructing physically relevant, rotating, nonextremal topological solitons.
We construct and analyze new smooth, horizonless, non-BPS geometries, which we call \emph{rotating topological stars}.
Unlike the solution in~\cite{Bianchi:2025uis}, these configurations include a Kaluza-Klein monopole (KKm) that ensures compatibility between smoothness at the cap and standard Kaluza-Klein asymptotics.
Following~\cite{Chakraborty:2025ger}, they are derived by applying a sequence of sigma-model transformations in five-dimensional minimal supergravity to the Kerr-Taub-bolt geometry.

The spacetime structure of the solutions is depicted in Fig.\ref{fig:Schematic}. We perform a detailed regularity analysis—covering smoothness at the cap and the absence of closed timelike curves~(CTCs)—and show that the solutions depend on a continuous parameter $-1<q<1$ and three integers $(k,\ell,N)$.
Here $k\geq0$ is the main quantum number controlling the angular momentum, $N$ is the quantized KKm charge, and $\ell$ encodes internal topological data.

These solitons carry all allowed charges in five-dimensional supergravity on S$^1$: electric and magnetic charges of the U(1) gauge field, $Q$ and $P$ (corresponding to M2- and M5-brane charges in M-theory on a rigid T$^6$), together with electric and magnetic charges of the Kaluza-Klein vector, $Q_0$ and $P_0$ (corresponding to P and KKm charges).
We analyze the tower of states labeled by $k$:
\begin{itemize}[noitemsep]\vspace{-0.3cm}
    \item $k=0$ gives static geometries that correspond to dyonic generalizations of the topological star of~\cite{Bah:2020ogh,Bah:2020pdz}.
    \item $k=1$ yields the BPS limit, corresponding to a spectral flow of a smooth $\frac12$-BPS Gibbons-Hawking center, the basic building block of BPS multicenter microstate geometries~\cite{Bena:2007kg,Heidmann:2017cxt,Bena:2017fvm,Warner:2019jll}.
    \item $k\geq2$ produces genuinely non-BPS rotating topological stars.
\end{itemize}

The $k\geq 2$ solutions possess an ergoregion, as rotating black holes do, but this exists only in five dimensions and disappears upon dimensional reduction to four dimensions.
This suggests an ergoregion instability~\cite{Friedman:1978ygc,1978RSPSA.364..211C,Cardoso:2005gj,Chowdhury:2007jx,Cardoso:2007az,Moschidis:2016zjy}  only for modes carrying momentum along the fifth dimension.

We also compute the gravitational multipole moments, showing that although their structure broadly mirrors that of charged STU black holes~\cite{Bena:2020uup}, important differences emerge.
In particular, the mass quadrupole moment, typically negative for black holes and most gravitating bodies, can be positive for rotating topological stars, such as the new solitons constructed in \cite{Bena:2025usu}. 
Such a spin-induced prolate shape—elongated along the rotation axis—contrasts sharply with the oblate profiles usually associated with rotating objects.\footnote{The quadrupole moment can be, in principle, extracted from the gravitational-wave signal of coalescing binaries~\cite{Krishnendu:2017shb}. Although with large measurement errors, gravitational-wave data do not yet exclude, and may even favor, positive mass quadrupole moments~\cite{LIGOScientific:2021sio}.} This unusual property, shared with the solutions of \cite{Bena:2025usu}, is discussed at length in that work, to which we refer the interested reader for a detailed analysis.

We further perform a numerical scan of the parameter space to identify regions where the rotating topological star shares the same mass, charges, and angular momentum as the generic STU black hole of~\cite{Chow:2014cca}.
Despite respecting the BPS bound, no such overlap is found for nonzero angular momentum, suggesting that these states lie outside the black-hole extremality bound and, unlike their static counterparts, do not coexist with black holes.

Nevertheless, we uncover a regime where the rotating topological star is nearly vacuum and its conserved charges closely approximate those of a Kerr black hole embedded in five dimensions with a large boost along the compact direction.
From a phenomenological standpoint, these configurations provide meaningful prototypes of coherent astrophysical black-hole microstates in string theory.

Finally, we show that rotating topological stars possess all the necessary ingredients for a tractable analysis of their dynamics under linear perturbations.
Both the geodesic equations for test particles and the Klein-Gordon equation for a minimally coupled scalar field separate into radial and angular components.
This property makes it possible to study photon scattering, photon rings, imaging simulations, scalar stability, quasinormal modes, and ringdown signals~\cite{Heidmann:2022ehn,Heidmann:2023ojf,Dima:2024cok,Bena:2024hoh,Dima:2025zot,Dima:2025tjz} (avoiding the technicalities of non-separable problems, see e.g.~\cite{Ikeda:2021uvc}), which we leave for future work.

Our results thus provide a concrete realization of smooth, rotating, horizonless objects in asymptotically flat spacetimes, extending earlier constructions in string theory and supergravity, and opening new directions for exploring the role of topological solitons in the gravitational phase space and the construction of rotating coherent microstates of nonextremal black holes in string theory. \\

Section~\ref{sec:rotating} introduces the new rotating topological stars and analyzes their geometry. Section~\ref{sec:TowerSol} discusses the tower of regular solutions and determines whether they coexist with a black hole at fixed conserved charges. In Section~\ref{sec:TSclosetoKerr}, we identify a region of the parameter space where the topological star approaches the boosted Kerr black string and analyze the corresponding solutions. Section~\ref{sec:GravSign} establishes the separability of the geodesic and scalar wave equations. Finally, Section~\ref{sec:discussion} provides concluding remarks and an outlook. Several appendices complement the main text. Appendix~\ref{sec:SigmaModel5d} summarizes the sigma model governing cohomogeneity-two solutions in minimal supergravity, and its associated solution-generating techniques~\cite{Chakraborty:2025ger}. Appendix~\ref{sec:Kerr-Taub-bolt} reviews the Kerr-Taub-bolt seed geometry, emphasizing the conditions required to reconcile regularity at the cap with the correct Kaluza-Klein asymptotics. Appendix~\ref{sec:ConstructionSteps} details the sigma-model transformations used to generate the rotating topological star from the Kerr-Taub-bolt geometry. \\

This work was developed in close discussion with Iosif Bena and Angèle Lochet, who were independently constructing five-dimensional rotating geometries using the ``running bolt'' method of adding BPS flux to the Kerr-Taub bolt \cite{Bena:2009qv}. Our paper and theirs \cite{Bena:2025usu} appeared simultaneously.

\section{Asymptotically-flat rotating topological stars} \label{sec:rotating}

The geometries constructed in this paper are solutions of five-dimensional $\cN=2$ minimal supergravity, determined by the action: 
\begin{equation}
\cS_5 \=  \frac{1}{16 \pi G_5} \int \left( R_5 \star_5 1 
\!-\! \frac32 \,F \!\wedge\! \star_5 F \!-\! F \!\wedge\! F \!\wedge\! A \right),
\label{eq:L5min}
\end{equation}
where $G_5$ is the Newton constant, $F = dA$ is the field strength of the gauge field $A$, $R_5$ is the Ricci scalar, and $\star_5$ is the Hodge star operator, defined with respect to the five-dimensional metric. This theory can be trivially embedded in M-theory on a rigid $T^6$, and it also arises from the consistent truncation of $\mathcal{N}=2$ five-dimensional supergravity coupled to two vector multiplets, with all multiplets identified trivially.

\subsection{Generating rotating solitons from Kerr-Taub-bolt}

A new roadmap for constructing non-BPS smooth geometries in five-dimensional supergravity was recently developed in~\cite{Chakraborty:2025ger}. A review of this method is provided in Appendix~\ref{sec:SigmaModel5d}. The approach relies on reducing the five-dimensional action to an integrable three-dimensional sigma model, whose hidden symmetries allow one to generate new non-BPS solutions by applying algebraic transformations to a seed geometry. This produces families of solutions that would be prohibitively difficult to obtain by directly solving the supergravity equations. This technique has already been successfully used to generate the most general nonextremal black hole solutions in STU supergravity, starting from the Kerr-NUT geometry as the seed~\cite{Cveti__1996,Chow:2014cca}. Strikingly, it has never been applied to construct smooth horizonless geometries using a smooth seed rather than a black hole, which is precisely the program initiated in~\cite{Chakraborty:2025ger}.

The first horizonless geometries obtained with this strategy, the static $\cW$-solitons~\cite{Chakraborty:2025ger,Dima:2025tjz}, demonstrate the feasibility of the approach using the Taub-bolt geometry as a seed. Here, we go one step forward by applying the same strategy to a seed solution allowing rotation after transformation: the Kerr-Taub bolt. This is a vacuum solution of five-dimensional gravity, obtained by Wick rotating the Kerr-NUT spacetime and adding a trivial time direction. The Kerr-Taub-bolt metric depends on three parameters $(m,n,a)$, is asymptotic to $\IR^{1,3}\times$S$^1$, and is given by
\begin{align}
    ds^2 &\= -dt^2 + \frac{\Delta +a^2 \sin^2 \theta}{\Sigma} \left(d\psi + \omega_\psi\, d\phi \right)^2 + \frac{\Sigma}{\Delta +a^2 \sin^2 \theta} \,ds_3^2\,, \nn\\
    ds_3^2 &\= \left(\Delta +a^2 \sin^2 \theta \right) \left(\frac{dr^2}{\Delta} + d\theta^2 \right) + \Delta \sin^2 \theta \,d\phi^2,\label{eq:KerrTaubBolt}
\end{align}
where
\begin{align}
    \Delta &\equi r^2-2mr+n^2-a^2\,,\qquad R \equi n(r-m)-a m \cos \theta\,, \nn\\
    \omega_\psi &\= 2\left[n (\cos \theta+1)-a \sin^2 \theta \, \frac{m R +(m^2-n^2)(n+a \cos \theta)}{n (\Delta +a^2 \sin^2 \theta)} \right]\,,\label{eq:KerrTaubBoltFields}\\
    \Sigma & \equi \Delta + a^2 \sin^2 \theta + \frac{2}{n} \left[ m R +(m^2-n^2)(n+a \cos \theta)\right] \,. \nn
\end{align}
The geometry caps off at a bolt located at $r=r_+$, the largest of the two roots of $\Delta$:
\begin{equation}
    r_\pm \equi m \pm \sqrt{m^2+a^2-n^2}\,.
    \label{eq:rpmDef}
\end{equation}
The parameters have the following interpretation: $m$ is associated with the ADM mass, $n$ with the Kaluza-Klein monopole charge, and $a$ with the KK dipole moment. A more convenient parametrization uses $(r_+,r_-,a)$:
\begin{equation}
    m \= \frac{r_++r_-}{2}\,,\qquad n \= \pm \sqrt{a^2 + r_- r_+}\,.
    \label{eq:m&nrpm}
\end{equation}

Some sigma-model transformations reshuffle the supergravity fields while preserving the structure of the seed spacetime, including its bolt smoothness and asymptotics. Thus, $m$ continues to produce mass (but also charges), and $a$ can generate spin after transformation. The role of the KK monopole charge $n$, however, deserves emphasis. Far from being superfluous, $n \neq 0$ is in fact essential: the Kerr-bolt geometry ($n=0$) cannot be simultaneously smooth at the bolt and asymptotic to $\IR^{1,3}\times$S$^1$~\cite{Dowker:1995gb}. A detailed analysis of this compatibility and the Kerr-Taub-bolt structure is provided in Appendix~\ref{sec:Kerr-Taub-bolt}.

Starting with a seed carrying nonzero KK monopole charge is therefore crucial for generating   rotating geometries via sigma-model transformations that are both smooth in the interior and asymptotic to a realistic Kaluza-Klein space: a trivial S$^1$ over a four-dimensional Minkowski spacetime. By contrast, a recent work~\cite{Bianchi:2025uis} constructed a ``rotating topological star'' from the Kerr-bolt geometry ($n=0$). While the solution has many interesting features, the smoothness in the interior imposes unrealistic asymptotics $(\IR^{1,3}\times$S$^1)/\mathbb{Z}_q$ (see~\cite{Dowker:1995gb} and Appendix~\ref{sec:CondAsympSmooth}), corresponding to a S$^1$ fibered over a \emph{Melvin universe}, also called \emph{magnetic flux tube}~\cite{Melvin:1963qx}, different from our Minkowski spacetime.

Here, by contrast, we construct rotating topological stars via carefully chosen sigma-model transformations on the Kerr-Taub-bolt geometry, ensuring standard Kaluza-Klein asymptotics through the global identifications of the compact directions:
\begin{equation}
(\phi,\psi) \= (\phi,\psi) + (2\pi,0)\,,\qquad (\phi,\psi) \= (\phi,\psi) + (0,2\pi R_\psi)\,,
\label{eq:AngleIdentification}
\end{equation}
with $R_\psi$ the radius of the S$^1$. The construction steps, including the sigma-model transformations used, are detailed in Appendix~\ref{sec:ConstructionSteps}; here we simply present the final result.

\subsection{The rotating topological star}
\label{sec:TheSol}

Our rotating topological star is intrinsically more complicated than the static topological star pioneered in~\cite{Bah:2020ogh,Bah:2020pdz}. 
Compared to the latter, it involves the dipole parameter $a$, but also a charge parameter $n$ and a residual transformation parameter $q\in(-1,1)$, which enables the conversion of the KK dipole $a$ into a spin. The metric and gauge field take the form
\begin{align}
ds_5^2 \= &\frac{Z F_1-F_2^2}{Z^2}\,\left[d\psi + \frac{Z F_3-F_2 F_4}{Z F_1-F_2^2}\,(dt+\omega_t d\phi)+ \omega_\psi d\phi \right]^2+ \frac{Z}{\sqrt{Z F_1-F_2^2}} \,ds_4^2 \nn\\
A \= &  \frac{F_4}{Z} \,(dt+\omega_t d\phi) + \frac{F_2}{Z} \,(d\psi+\omega_\psi d\phi) \label{eq:RTSMetric&Field}\\
&+ \frac{q \left({m^{-}_{1,6}}^2-4 q^6 a^2 \right)}{(1-q^2){m^{+}_{1,6}}^2}  \left[(r_++r_-)(\cos\theta+1) + \frac{2a R \sin^2\theta}{\Delta + a^2 \sin^2\theta}   \right]\,d\phi\,, \nn
\end{align}
where $ds_4^2$ is given by
\begin{equation}
ds_4^2 \= - \frac{\Delta+a^2 \sin^2 \theta}{\sqrt{Z F_1-F_2^2}}\,(dt+\omega_t d\phi )^2 + \sqrt{Z F_1-F_2^2} \left( \frac{dr^2}{\Delta} +d\theta^2 + \frac{\Delta}{\Delta+a^2 \sin^2 \theta}\,\sin^2 \theta \, d\phi^2\right),
\label{eq:4dMetricTS}
\end{equation}
and the fields are given by
\begin{align}
F_1 &\equi  \Delta + a^2 \sin^2 \theta + \frac{q^2(r_++r_-)}{ n(1-q^2)  m^{-}_{2,6} {m^{+}_{1,6}}^2}  \Biggl\{ m^{-}_{2,4} {m^{+}_{1,6}}^2R  +4 q^4 n^2 \left[\frac{2(1-q^2)r_+ r_-}{r_++r_-}{m^{+}_{1,6}}-{m^{-}_{2,8}}  \right]R  \nn \\
&\qquad \qquad  \quad + \frac{m^{-}_{1,0}{}^{2}-4 a^2}{2(r_++r_-)} \left[ {m^{+}_{1,6}}^2 \left(a m^{-}_{2,4}  \cos\theta  - n m^{+}_{2,4} \right)+4 q^4 n^2 \left(n m^{+}_{2,8} - a m^{-}_{2,8}  \cos\theta  \right) \right] \Biggr\},  \nn \\
F_2 &\equi - \frac{q \sqrt{{m^{-}_{1,6}}^2-4 q^6 a^2}}{(1-q^2)^\frac{3}{2} {m^{+}_{1,6}}^2 \sqrt{m^{-}_{2,6}}} \left\{2(1-q^4)m^{-}_{2,6} \,R + q^4 \left[ m^{-}_{1,0}{}^{2}-4 a^2\right]  \left(n m^{-}_{1,2} -a m^{+}_{1,2} \cos\theta  \right)   \right\}, \nn  \\
F_3 &\equi \frac{q^3 r_- r_+}{n m^{-}_{2,6} {m^{+}_{1,6}}^2}  \Biggl\{ R \, \left[ (r_++r_-)\left({m^{+}_{1,6}}^2+4q^6n^2 \right) -4 {m^{+}_{1,6}} {m^{+}_{2,6}} \left(1+\frac{a^2}{r_+ r_-} \right) \right]\label{eq:FieldsRTS1}\\
&\qquad \qquad \qquad + \frac{m^{-}_{1,0}{}^{2}-4 a^2}{2} \left[ {m^{+}_{1,6}}^2 \left(a \cos \theta +  \frac{1+q^2}{1-q^2}n  \right) +4 q^6 n^2 \left(a \cos \theta -  \frac{1+q^2}{1-q^2} n  \right) \right] \Biggr\} , \nn \\
F_4 &\equi \frac{q^2 \sqrt{{m^{-}_{1,6}}^2-4 q^6 a^2} }{(1-q^2)^\frac{3}{2} {m^{+}_{1,6}}^2 \sqrt{m^{-}_{2,6}}}  \left\{ 2(1-q^2) m^{-}_{2,6}\, R -q^2 \left[ m^{-}_{1,0}{}^{2}-4a^2\right] \left( m^{+}_{1,4} a \cos\theta  + n m^{-}_{1,4}\right) \right\},  \nn \\
Z &\equi  \Delta + a^2 \sin^2 \theta + \frac{(1+q^2) \,\left({m^{-}_{1,6}}^2-4 q^6 a^2 \right)}{n(1-q^2) {m^{+}_{1,6}}^2}  \left[ m^{+}_{1,0} R  +  \frac{ m^{-}_{1,0}{}^{2}-4a^2 }{2} \left( a \cos \theta + \frac{1+q^4}{1-q^4} n \right) \right] ,  \nn \\
\omega_t &\=  \frac{a q^3 \left( m^{-}_{1,0}{}^{2}-4a^2 \right)  \sqrt{{m^{-}_{1,6}}^2-4 q^6 a^2} }{(1-q^2)^\frac{3}{2} m^{+}_{1,6}\sqrt{m^{-}_{2,6}} }\,\frac{(r - r_+) \,\sin^2 \theta}{\Delta + a^2 \sin^2\theta}, \nn\\
\omega_\psi &\=  \frac{  \sqrt{{m^{-}_{1,6}}^2-4 q^6 a^2}  }{ 2(1-q^2)^\frac{3}{2} \, {m^{+}_{1,6}}^2 \sqrt{m^{-}_{2,6}}  }  \Biggl\{ 4n \left(1 -q^6\right)m^{-}_{2,6} (1+\cos \theta) +  \frac{a \sin^2\theta}{n(\Delta+a^2\sin^2 \theta)} \nn \\
&\qquad  \times \left[2(r_++r_-) \left({m^{-}_{1,6}}^2-4 q^6 a^2 \right) \,R + \left( m^{-}_{1,0}{}^2-4a^2 \right) \left( a {m^{+}_{1,6}}^2 \cos \theta + n {m^{-}_{2,12}}   \right) \right]  \Biggr\} . \nn
\end{align}
Here $\Delta$, $R$ and $n$ are given by the Kerr-Taub-bolt values~\eqref{eq:KerrTaubBoltFields} and~\eqref{eq:m&nrpm}, and we have defined 
\begin{align}
m^{\pm}_{i,j} \equiv {r_+}^i \pm q^j \,{r_-}^i .
\end{align}
One can verify that setting $q=0$ (no transformation) reproduces the Kerr-Taub-bolt solution~\eqref{eq:KerrTaubBolt}, while taking $a=n=r_-=0$ yields the static topological star of~\cite{Bah:2020ogh,Bah:2020pdz}, with $(r_\text{B},r_\text{S})=\tfrac{r_+}{1-q^2}(1,q^2)$.

\subsection{Conserved charges and asymptotics}
\label{sec:ConservedCharges}

Provided the identifications~\eqref{eq:AngleIdentification} are satisfied, the solution asymptotes to $\mathbb{R}^{1,3} \times S^1$. Upon KK reduction along $\psi$, the four-dimensional metric is given by~\eqref{eq:4dMetricTS}. The solution describes a rotating configuration with four charges: electric and magnetic components for both the Kaluza-Klein and U(1) gauge fields. The conserved charges are
\begin{align}
P_0  &\=  \frac{2n \left(1 -q^6\right) \sqrt{m^{-}_{2,6}\left({m^{-}_{1,6}}^2-4 q^6 a^2\right)}  }{ (1-q^2)^\frac{3}{2} \, {m^{+}_{1,6}}^2  } ,\qquad P \=  -\frac{q(r_++r_-) \left({m^{-}_{1,6}}^2-4 q^6 a^2 \right)}{(1-q^2){m^{+}_{1,6}}^2} ,\nn\\
Q_0 &\= \frac{q^3 r_- r_+}{m^{-}_{2,6} {m^{+}_{1,6}}^2}  \, \left[ (r_++r_-)\left({m^{+}_{1,6}}^2+4q^6n^2 \right) -4 {m^{+}_{1,6}} {m^{+}_{2,6}} \left(1+\frac{a^2}{r_+ r_-} \right) \right] ,\nn\\
Q & \= \frac{-2 n \, q^2\sqrt{m^{-}_{2,6}\left({m^{-}_{1,6}}^2-4 q^6 a^2\right)} }{{m^{+}_{1,6}}^2 \sqrt{1-q^2}}  ,  \quad J \= \frac{a q^3 \left( m^{-}_{1,0}{}^{2}-4a^2 \right)  \sqrt{{m^{-}_{1,6}}^2-4 q^6 a^2} }{2(1-q^2)^\frac{3}{2} m^{+}_{1,6}\sqrt{m^{-}_{2,6}} } , \nn \\
M& \= \frac{(r_++r_-)}{4 (1-q^2)  {m^{+}_{1,6}}^2 }  \Biggl\{ \frac{q^2}{m^{-}_{2,6}}\left[ m^{-}_{2,4} {m^{+}_{1,6}}^2  +\frac{8 q^4 n^2(1-q^2)r_+ r_-}{r_++r_-}{m^{+}_{1,6}}-4 q^4 n^2{m^{-}_{2,8}}\right]  \nn \\
& \hspace{3.5cm} + (1+q^2) \,\left({m^{-}_{1,6}}^2-4 q^6 a^2 \right) \Biggr\}, \label{eq:ConservedCharges}
\end{align}
where the index ``0" indicates the KK charges,  $P$ and $Q$ are for magnetic and electric respectively,  and $M$ and $J$ denote the mass and spin, respectively, in units where $G_4=1$.  

There are several special limits of interest:
\begin{itemize}
\item[•] \underline{The vacuum limits}: 

There are, in principle, two exact vacuum limits: $q = 0$ and ${m^{-}_{1,6}}^2 - 4 q^6 a^2 = 0$. The first simply undoes the transformations, leading back to the Kerr-Taub-bolt geometry. The second yields non-trivial values for $M$ and $Q_0$, but a careful analysis reveals that $M$ is negative, rendering the solution unphysical.

However, there exists a ``non-exact'' vacuum limit in which $m^{-}_{2,6}$ approaches zero:
\begin{equation}
m^{-}_{2,6} = r_+^2 \- q^6 \, r_-^2 \,\sim\,0.
\label{eq:VacuumLimit}
\end{equation}
This limit is non-exact because $m^{-}_{2,6}$ cannot be strictly zero as some quantities diverge. While it might appear challenging to achieve $m^{-}_{2,6} \sim 0$ given that $|q| < 1$ and $r_- < r_+$, we can freely choose $r_- < 0$ and $r_+ < -r_-$ so that $m^{-}_{2,6}$ can be made arbitrarily small.
In this regime, the mass, spin, and KK electric charge diverge, while $P_0$ and $Q$ vanish and $P$ remains finite. The solution is thus dominated by its vacuum components:\footnote{This behavior is reminiscent of certain black holes in the STU model~\cite{Chow:2014cca}, where some charges, as well as the mass and spin, can be made arbitrarily large by sending boost parameters to infinity without taking $m$ to zero. We will come back to this limit in more detail in Section~\ref{sec:TSclosetoKerr}.}
\begin{equation}
M\,,\, Q_0\,,\, J^2 \sim \cO\left(\frac{1}{m^{-}_{2,6}} \right) \quad \gg \quad P\sim \cO\left(1\right)\quad \gg \quad P_0\,,\,  Q  \sim \cO\left(m^{-}_{2,6}\right)\,.
\end{equation}
We will analyze these peculiar solutions in detail and show that this limit leads to perfectly physical rotating topological stars that are comparable to the vacuum boosted Kerr black string in Section \ref{sec:TSclosetoKerr}.

\item[•] \underline{The extremal limit}: 

As for the Kerr-Taub bolt~\eqref{eq:KerrTaubBolt}, the three-dimensional base becomes Ricci-flat when
\begin{equation}
r_+ - r_- \= 2|a|, \label{eq:ExtremalLimit}
\end{equation}
which corresponds to the extremal limit of the solution. In Section~\ref{sec:ExtremalLimit}, we will demonstrate that this extremal solution corresponds to the spectral flow of a smooth Gibbons-Hawking center that preserves 16 supercharges. Such a center serves as a fundamental building block for multicenter bubbling geometries, a class of microstates of BPS black holes in supergravity~\cite{Bena:2007kg,Heidmann:2017cxt,Bena:2017fvm,Warner:2019jll}.

\item[•] \underline{The static limit}: 

Although several of the limits discussed above lead to vanishing angular momentum ($J = 0$), the most natural static limit is achieved by setting $a = 0$. This corresponds to starting from the static limit of the Kerr-Taub-bolt solution, namely, the Taub-bolt geometry. In Section~\ref{sec:StaticSol}, we will show that these static solutions describe a four-charge extension of the topological star introduced in~\cite{Bah:2020ogh}.

\end{itemize}

\subsection{Regularity conditions}
\label{sec:RegCondTS}

The primary condition for regularity has already been imposed (see Appendix~\ref{sec:ConstructionSteps} for details): the coordinate singularity at $r = r_+$ corresponds to the degeneracy of a compact direction. To achieve this, we required that $\omega_t |_{r = r+} = 0$. Consequently, at $r = r_+$, the $\phi$ direction smoothly shrinks at fixed $\psi + \omega_\psi |_{r=r_+} \,\phi$.  

We now proceed to analyze the regularity conditions at the various coordinate degeneracies and ensure the absence of CTCs throughout the spacetime.

\subsubsection{At the coordinate degeneracies}

The solution has three types of coordinate degeneracies: at the poles of the $S^2$ (where $\theta = 0$ or $\pi$), and at the bolt locus $r = r_+$. We derive the regularity conditions at these loci and examine their intersections at $(r = r_+, \theta = 0)$, i.e. the North pole of the bolt, and $(r = r_+, \theta = \pi)$, i.e. the South pole of the bolt.

\begin{itemize}
\item[•] \underline{At the poles of the S$^2$:}
\end{itemize}

At $\theta = 0$ and $\pi$, we have $\omega_t |_{\theta = 0,\pi} = 0$, and $\omega_\psi |_{\theta = 0,\pi}$ is independent of $r$. Therefore, the shrinking direction is $\phi$, at fixed $\psi + \omega_\psi |_{\theta = 0,\pi} \phi$. Locally, the metric is $d\theta^2 + \sin^2 \theta \, d\phi^2$, and imposing the identifications~\eqref{eq:AngleIdentification} ensures a smooth $\mathbb{R}^2$ geometry if $\omega_\psi |_{\theta = 0,\pi}\in \mathbb{Z} R_\psi$. We found $\omega_\psi |_{\theta = 0}=2P_0$ and $\omega_\psi |_{\theta = \pi}=0$. Thus, regularity at the poles of the $S^2$ demands the quantization of the KKm charge:
\begin{equation}
P_0 \= \frac{1}{2} N R_\psi\,,\qquad N \in \mathbb{Z}\,.  \label{eq:KKmcharge}
\end{equation}

\begin{itemize}
\item[•] \underline{At the bolt:}
\end{itemize}

Next, we consider the locus $r = r_+$. Here, $\omega_\psi |_{r = r+}$ is constant and independent of $\theta$. Introducing the local radial coordinate $\rho$, 
\begin{equation}
\rho^2 \equi 4(r-r_+),
\label{eq:CoorBolt}
\end{equation}
and expanding near $\rho = 0$, the $(\rho, \phi, \psi)$ part of the metric becomes
\begin{equation}
d\rho^2 \+ \rho^2 \,\frac{m^{-}_{1,0}{}^{2}}{4a^2}\,d\phi^2 \+ \# \,\left( d\psi + \omega_\psi |_{r=r_+}\, d\phi\right)^2\,,
\label{eq:MetricBolt}
\end{equation}
where $\#$ corresponds to an irrelevant finite value. The shrinking direction is $\widetilde{\phi}$ at fixed $\widetilde{\psi}$ with
\begin{equation}
\widetilde{\phi} \equi \frac{r_+-r_-}{2|a|} \phi, \qquad \widetilde{\psi} \equi \psi+ \omega_\psi|_{r=r_+} \,\phi. \label{eq:LocalAngles}
\end{equation}
The global identifications~\eqref{eq:AngleIdentification} then yield the local periodicities 
\begin{equation}
\left(\widetilde{\phi},\widetilde{\psi} \right) \= \left(\widetilde{\phi},\widetilde{\psi} \right) + \begin{cases} 
2\pi \left( \frac{r_+-r_-}{2|a|} ,  \omega_\psi|_{r=r_+} \right)  \qquad \qquad &(A)\\
2\pi \left(0 , R_\psi \right) \qquad &(B)
\end{cases},
\end{equation}
The identifications leave $\widetilde{\psi}$ invariant when we take the combination $k \times$ (A) $- \ell \times$ (B) with
\begin{equation}
\omega_\psi|_{r=r_+} \= \frac{\ell}{k} \, R_\psi\,, \qquad (\ell,k) \in \mathbb{Z}\,,\qquad \gcd(\ell,k)=1,\qquad k\geq 0.
\label{eq:RegCondRpsi}
\end{equation}
A priori, the above arguments apply only for $a \neq 0$ and $k \geq 1$. However, as we will see shortly (and in more detail in Section~\ref{sec:StaticSol}), the regularity conditions for $k=0$ are also well defined and lead to the static limit $a=0$. We therefore extended the range to $k \geq 0$.
Smoothness at the bolt then requires the shift in $\phi$ under this combination, which is $2\pi \frac{r_+-r_-}{2|a|}k$, to be $2\pi$. This gives the condition:
\begin{equation}
\frac{r_+-r_-}{2|a|} k \= 1 \qquad \Leftrightarrow \qquad |a| \= \frac{k(r_+-r_-)}{2}\,.\label{eq:RegCond1}
\end{equation}
In terms of the parameters $(m, n, a)$, regularity at the bolt compatible with Kaluza-Klein asymptotics implies
\begin{equation}
    n \= \pm \sqrt{m^2+a^2\left(1-\frac{1}{k^2}\right)}\,.
    \label{eq:nFixKTB}
\end{equation}
This confirms our earlier claim: starting from a geometry with a nontrivial NUT charge $n$ is essential to obtain a smooth solution that is compatible with the standard Kaluza-Klein asymptotics defined by the global identification~\eqref{eq:AngleIdentification}. More strikingly, achieving this compatibility requires $|n|>m$ (for $k\geq 2$), meaning the seed solution lies outside the BPS bound (or equivalently the black hole bound), with an excess of charge relative to its mass (see Appendix~\ref{sec:Kerr-Taub-bolt} for more details). This will have a direct consequence: the resulting rotating topological star, as we will see later, will lie inside the BPS bound but outside the black hole extremality bound. \\

To summarize, imposing regularity at the bolt, together with the quantization of the KKm charge~\eqref{eq:KKmcharge}, introduces three integers $(k, \ell, N) \in \mathbb{Z}$, with $k \geq 0$, such that
\begin{equation}
    |a| \= \frac{k\,(r_+ - r_-)}{2},\qquad \omega_\psi|_{r=r_+} \= \frac{\ell}{k} \, R_\psi,\qquad P_0 \= \frac{1}{2} N R_\psi. \label{eq:RegCond}
\end{equation}
The integer $k$ can be interpreted as the \emph{primary quantum number}, as it discretizes the spin parameter $a$ into a tower of allowed values. The case $k = 0$ corresponds to the static limit $a = 0$.\footnote{Although~\eqref{eq:RegCond} appears ill-defined at $k = 0$, the quantities remain finite because $\omega_\psi |_{r = r+}$ scales as $a^{-1}$ so the $k^{-1}$ drops after replacing $a$ by $\pm k \frac{r_+-r_-}{2}$. A more detailed analysis of the static $a=0$ solutions will be given in Section~\ref{sec:StaticSol}.} The case $k = 1$ corresponds to the extremal limit~\eqref{eq:ExtremalLimit}, while $k \geq 2$ describes generic non-BPS rotating geometries terminating in a smooth bolt.

After some algebra, the last two conditions give
\begin{equation}
    R_\psi = \frac{4(1-q^6)\sqrt{m^{-}_{2,6}(a^2+r_+r_-)\left({m^{-}_{1,6}}^2-4a^2 q^6 \right)}}{N (1-q^2)^\frac{3}{2} {m^{+}_{1,6}}^2},\quad \frac{2\ell}{k N} = 1 + \frac{(r_+-r_-)m^{-}_{1,6}-2a^2(1+q^6)}{2 a(1-q^6) \sqrt{a^2+r_- r_+}}\,. \label{eq:RegQuantization}
\end{equation}
Remarkably, the second condition can be inverted to yield the quantization of $r_-$, together with that of $a$ from~\eqref{eq:RegCond}, in terms of the three integers $(k, \ell, N)$:
\begin{equation}
    r_- \= \frac{
\frac{k^2 (1 + q^6)^2}{4 q^6}
-
\left[
\frac{(1 - q^6)(k N - 2 \ell)}{2 N |q|^3}
+
\epsilon \sqrt{
1 - \frac{\ell (\ell - k N)(1 - q^6)^2}{(k^2-1) N^2 q^6}
}
\right]^2
}
{
\left[ k^2 - 1 + \frac{k \ell (1 - q^6)}{N q^6} \right]
\left[ 1 + \frac{(N - k \ell)(1 - q^6)}{N (k^2 - 1)} \right]
}\,r_+,\qquad a= \epsilon  \frac{k\,(r_+ - r_-)}{2}\,,
\label{eq:RmQuantization}
\end{equation}
where $\epsilon = \pm 1$ fixes the sign of $a$ (and we consider $\epsilon=1$ for $a=0$). Finally, $r_+$ can be expressed in terms of the extra-dimension radius $R_\psi$, the transformation parameter $q$, and the three integers using the first equation in~\eqref{eq:RegQuantization}. \\

Overall, solutions that are smooth at the bolt and compatible with S$^1 \times \IR^{1,3}$ asymptotics are parametrized by a continuous parameter $q\in(-1,1)$ and three integers $(k, \ell, N)$, with $R_\psi$ fixed by the theory. The spectrum of states is intricate, and the range of $(k, \ell, N)$ for which solutions exist is nontrivial. In Sections~\ref{sec:TowerSol} and \ref{sec:TSclosetoKerr}, we will analyze in detail the tower of solutions labeled by the primary number $k$, and show that, unlike the topological star~\cite{Bah:2020ogh}, there exist solutions for which $(r_+, r_-, a)$ can be taken large compared to the KK scale $R_\psi$.

\begin{itemize}
\item[•] \underline{At the poles of the bolt:}
\end{itemize}

At the poles of the bolt, the local spherical coordinates are
\begin{equation}
\begin{split}
r &\= \frac{1}{2}\left(r_{N/S} +r_++r_- + \sqrt{(r_{N/S}\cos \theta_{N/S}+r_+-r_-)^2+r_{N/S}^2 \sin^2 \theta_{N/S}} \right) \\
\cos \theta &\=   \pm \frac{\sqrt{(r_{N/S}\cos \theta_{N/S}+r_+-r_-)^2+r_{N/S}^2 \sin^2 \theta_{N/S}} -r_{N/S}}{r_+-r_-}\,,
\label{eq:NorthSouthCoor}
\end{split}
\end{equation}
where ``$N$'' stands for north pole with a ``$+$'',  while ``$S$'' stands for south pole with a ``$-$''.

After some algebra,  we find that the $(r_S,\theta_S,\phi,\psi)$ space at $r_S \to 0$ is given by
\begin{equation}
\frac{dr_S^2}{r_S} + r_S\, d\theta_S^2 + 4 r_S \sin^2 \frac{\theta_S}{2} \left(d\phi + \frac{k}{\ell R_\psi} d\psi \right)^2 +4 r_S \cos^2 \frac{\theta_S}{2} \frac{d\psi^2 }{\ell^2 R_\psi^2},
\end{equation}
which describes a regular $\mathbb{R}^4 / \mathbb{Z}_{|\ell|}$ space after introducing the coordinates $\rho_S = r_S^2$, $2 \tau_S = \theta_S$, $\varphi_1 = \phi + \frac{k}{\ell R\psi} \psi$, and $\varphi_2 = \frac{1}{\ell R_\psi} \psi$. The south pole thus corresponds to a NUT center with orbifold parameter $|\ell|$.

At the north pole, we similarly find
\begin{equation}
\begin{split}
\frac{dr_N^2}{r_N} + r_N\, d\theta_N^2 &+ 4 r_N \sin^2 \frac{\theta_N}{2} \left(\frac{\ell}{\ell-N k}\,d\phi +\frac{k}{\ell-N k}\, \frac{d\psi}{R_\psi} \right)^2  \\
&+4 r_N \cos^2 \frac{\theta_N}{2} \left(\frac{N}{\ell-N k}\,d\phi +\frac{1}{\ell-N k}\, \frac{d\psi}{R_\psi} \right)^2.
\end{split}
\end{equation}
Note that $\ell \neq N k$ since $\gcd(\ell,k)=1$, except if $k=1$ which will be treated separately.  As argued in~\cite{Bossard:2017vii}, this corresponds to a smooth $\mathbb{Z}_{|\ell-N k|}$ quotient of $\IR^4$.

\subsubsection{Regularity elsewhere}

Regularity at the bolt is not the only criterion for a physically admissible solution. One must also identify a regime of parameters for which the transformed geometries, while regular at the bolt, satisfy the following conditions:
\begin{itemize}
\item \underline{Absence of signature change in spacetime:} A closer examination of the Kerr-Taub-bolt metric~\eqref{eq:KerrTaubBolt} shows that the geometry changes signature wherever $\Sigma$ changes sign. As discussed in Appendix~\ref{sec:RegKerrTaubBolt}, this occurs for $r \geq r_+$ if $|n| > m$, which is precisely the regime imposed by regularity for topological stars with $k \geq 2$~\eqref{eq:nFixKTB}.

Consequently, one must ensure that such a signature change, which would also induce singular points, does not occur for the rotating topological star. This requires that $Z F_1 - F_2^2 > 0$ and $Z > 0$ everywhere.
\item \underline{Absence of CTCs:} A sufficient condition is that $t$ defines a global time function, which requires $g^{tt} < 0$ everywhere. We have:
\begin{equation}
    g^{tt} = - \frac{Z F_1 - F_2^2}{Z(\Delta + a^2 \sin^2 \theta)} + \frac{(\Delta + a^2 \sin^2 \theta)\, \omega_t^2}{Z \Delta \sin^2 \theta}.
\end{equation}
\item \underline{Positive mass:} The solution must satisfy $M > 0$ where $M$ is given in~\eqref{eq:ConservedCharges}.
\end{itemize}

After a detailed analysis of these three conditions for different values of the primary quantum number $k$, we have found that, for each $k$, there exists a range of parameters for which all three conditions are satisfied. These regimes are:
\begin{equation}
    \begin{split}
        &k=0 \text{ regular static solutions} \quad \Leftrightarrow \quad \frac{r_-}{r_+} \geq 0,\\
        &k=1 \text{ regular extremal solutions} \quad \Leftrightarrow \quad \frac{r_-}{r_+} \geq -1 \,\,\,\,\text{or}\,\,\,\, \frac{r_-}{r_+} < -|q|^{-3}, \\
        &k\geq 2 \text{ regular rotating solutions}  \quad \Leftrightarrow \quad \frac{r_-}{r_+} < -|q|^{-3}, 
    \end{split}
    \label{eq:RangeParam2}
\end{equation}

Note that, a priori, $(r_-, r_+)$ are not free parameters, as they are constrained by the quantization conditions~\eqref{eq:RegQuantization} and~\eqref{eq:RmQuantization} in terms of the three integers $(k, \ell, N)$. However, many values of these integers yield solutions within the three ranges of $r_-/r_+$ described above. The corresponding classes of solutions for the three values of $k$ will be studied in detail in Section~\ref{sec:TowerSol}.

Furthermore, it is evident that for $k \geq 2$, regularity requires $q \neq 0$. This means that the Kerr-Taub-bolt geometry ($q=0$) cannot meet all regularity conditions (it has regions of signature change). It is therefore remarkable that the sigma-model transformation opens up a region of parameter space in which the ambipolarity of the seed geometry is resolved.

Finally, the regime of validity for $k \geq 2$ solutions includes the near-vacuum limit $m^{-}_{2,6} \sim 0$, corresponding to $\frac{r-}{r_+} \sim -|q|^{-3}$. We will explore this specific limit in Section~\ref{sec:TSclosetoKerr}, and show that in this regime, the smooth horizonless geometries closely resemble a \emph{highly boosted Kerr black hole.}

\subsection{Ergoregions}

We now examine the possibility that our solutions contain ergoregions. We distinguish two types: a five-dimensional ergoregion, which occurs where
\begin{equation}
g_{tt}^{(5d)} \= \frac{(F_2 F_4-F_3 Z)^2-(\Delta+a^2 \sin^2 \theta) Z^3}{Z^2(Z F_1-F_2^2)} \,>\,0\,,
\end{equation}
and a four-dimensional ergoregion, which requires
\begin{equation}
g_{tt}^{(4d)} \= - \frac{\Delta +a^2 \sin^2 \theta}{\sqrt{\cI_4} } \,>\,0 \,.
\end{equation}

It is clear that if a solution is regular, then $g_{tt}^{(4d)} < 0$ everywhere, and no four-dimensional ergoregion can exist. In contrast, STU black holes typically have
$g_{tt}^{(4d)} = - \frac{\Delta -a^2 \sin^2 \theta}{\sqrt{\cI_4} } $ which necessarily changes sign above the horizon, where $\Delta=0$, thereby defining an ergoregion. The absence of a four-dimensional ergoregion in topological stars is therefore mainly due to the analytic continuation $a \to i a$, required by the Wick rotation from a black seed (Kerr–NUT) to a smooth seed (Kerr-Taub bolt).

However, we find that regular rotating topological stars with $k\geq 2$, satisfying~\eqref{eq:RangeParam2}, have an ergoregion localized near the cap in five dimensions. The fact that solutions possess an ergoregion in five dimensions but not in four suggests an ergoregion instability~\cite{Friedman:1978ygc,1978RSPSA.364..211C,Cardoso:2005gj,Chowdhury:2007jx,Cardoso:2007az,Moschidis:2016zjy} for modes carrying momentum along the fifth dimension. Since this momentum is quantized in units of $R_\psi^{-1}$, such modes correspond to high-energy excitations from a four-dimensional perspective if $R_\psi$ is only slightly larger than the Planck length.

\subsection{Multipole moments}
\label{sec:Multipoles}

As for any rotating gravitational solution, the geometry is characterized by a sequence of mass and spin multipole moments, which we derive in this section. We follow the method of Thorne~\cite{Thorne:1980ru}, as reviewed in~\cite{Bena:2020uup}. We begin by introducing the ``asymptotically Cartesian'' coordinates, AC-$\infty$, defined as
\begin{equation}
    r_s \sin \theta_s \= \sqrt{r^2-a^2} \sin \theta,\qquad r_s \cos \theta_S \= r \cos \theta\,.
\end{equation}
These coordinates are closely related to the AC-$\infty$ coordinates for Kerr(-Newman), differing by the substitution $a \to i a$, reflecting again the analytic continuation of the spin parameter relating both solutions.

Using the prescription of~\cite{Bena:2020uup}, we compute the ``raw moments'' $\widetilde{M}_p$ and $\widetilde{S}_p$:\footnote{Note that $\widetilde{M}_p$ and $\widetilde{S}_p$ are not the physical multipole moments since they are not gauge invariant. This is clearly shown by a nonzero $\widetilde{M}_1$.}
\begin{equation}
    \widetilde{M}_{2p} \= M \, a^{2p}, \qquad \widetilde{M}_{2p+1} \= \overline{M} \, a^{2p+1},\qquad \widetilde{S}_{2p} \= 0 \,,\qquad \widetilde{S}_{2p+1} \= J \,a^{2p}, \label{eq:multipoles}
\end{equation}
where $M$ and $J$ are the ADM mass and angular momentum of the solution, and
\begin{equation}
    \overline{M}\= \frac{-\sqrt{a^2+r_- r_+}\,\left[m^{-}_{1,6} {m^{-}_{1,4}}^2 \left(m^{+}_{1,6}+2 q^2 m_{1,2}^{+}\right) -4 a^2 q^6 \left(2 m^{-}_{1,8}+q^2 m^{-}_{1,4} \right) \right] }{2(1-q^2) {m^{+}_{1,6}}^2 m^{-}_{2,6}}.
\end{equation}
Since $\widetilde{M}_1 \neq 0$, this coordinate system is not ``asymptotically Cartesian and mass-centered'' (ACMC-$\infty$). The ACMC system is obtained by shifting $r_s \cos \theta_s \to r_s \cos \theta_s + \frac{\overline{M}}{M}$ while keeping $r_s \sin \theta_s$ unchanged. This yields the gauge-invariant mass and spin multipoles:
\begin{align}
 \label{eq:truemultipoles} M_p &= \sum_{j=0}^p \left(\begin{array}{c} p\\ j\end{array}\right)  \widetilde{M}_j \left(-\frac{a \overline{M}}{M}\right)^{p-j}, & S_p & = \sum_{j=0}^p \left(\begin{array}{c} p\\ j\end{array}\right)  \widetilde{S}_j \left(-\frac{a \overline{M}}{M}\right)^{p-j}.
\end{align}
These polynomials can be simplified and give:
\begin{equation}
\begin{split}
    M_p &\= \frac{1}{2} \left(-\frac{a}{M} \right)^p\,(M^2-{{\overline{M}}}^2)\left(({{\overline{M}}}+M)^{p-1}-({{\overline{M}}}-M)^{p-1} \right),\\
    S_p &\= \frac{1}{2} \left(-\frac{a}{M} \right)^{p-1}\,\frac{J}{M}\,\left(({{\overline{M}}}+M)^{p}-({{\overline{M}}}-M)^{p} \right).
\end{split}
\label{eq:MultipolesTS}
\end{equation}
Remarkably, this multipole structure matches that of the STU black hole~\cite{Bena:2020uup} upon substituting $\overline{M} \to \frac{\widetilde{M}_1}{a}$ and complexifying $a \to i a$.

This complexification affects the sign of the moments. For example, consider the mass and spin quadrupoles for the topological star $(M_2, S_2)$ and the STU black hole $(M_2^{\rm BH}, S_2^{\rm BH})$:\footnote{While $M$ can be considered identical for both the topological star and the black hole, $\overline{M}$ and $a$ does not need to be.}
\begin{align}
    M_2 &\= \frac{a^2(M^2-{{\overline{M}}}^2)}{M},\;\;\qquad S_2 \= -\frac{2a J {{\overline{M}}}}{M},\\ M^\text{BH}_2 &\= -\frac{a^2(M^2+{{\overline{M}}}^2)}{M},\quad S^\text{BH}_2 \= -\frac{2a J {{\overline{M}}}}{M}.
\end{align}
Thus, the mass quadrupole of the black hole is necessarily negative (as for Kerr, where $\overline{M}=0$), whereas for the topological star it can be either positive or negative. In addition, unlike Kerr, rotating topological stars and the generic STU black holes of \cite{Chow:2014cca} break equatorial symmetry and have a nonvanishing quadrupole spin moment. 

A positive quadrupole induced purely by rotation (since $M_2 = 0$ if $a = J = 0$) is highly exotic:\footnote{Within General Relativity, a positive quadrupole moment has been observed in certain gravastar models at large compactness~\cite{Uchikata:2016qku}.} it implies a prolate shape, elongated along the rotation axis, in contrast to the oblate shapes typical of rotating bodies such as black holes.
Measuring the quadrupole moment is in principle possible with the gravitational-wave signals from coalescing binaries~\cite{Krishnendu:2017shb}. Although measurement errors are still large, it is interesting to note that gravitational-wave data do not yet exclude positive quadrupole moments, and could even slightly favor them~\cite{LIGOScientific:2021sio}.

However, the sign of $M_2$ crucially depends on whether $M \lessgtr |\overline{M}|$, which is parameter-dependent. For the nonextremal rotating topological stars with $k\geq 2$ in the validity range~\eqref{eq:RangeParam2}, we found that the quadrupole moment is negative like a black hole, except in a small window close to 
\begin{equation}
    r_- \,\sim\, -\frac{\sqrt{2+q^2}}{|q|^3 \sqrt{1+2q^2}}\,r_+\,.
    \label{eq:RegionOfPositiveM2}
\end{equation} 
The possibility of a positive mass quadrupole might therefore be a characteristic feature of these non-supersymmetric smooth horizonless geometries. In \cite{Bena:2025usu}, it is conjectured that this behavior originates from the ambipolarity of the four-dimensional base space, which effectively acts as a ``negative mass source'' in spacetime, even though the total mass remains positive. Because a negative-mass region is ``lighter'' than the vacuum, the rotating configuration becomes prolate—just as a lighter fluid forms an elongated shape when spun inside a heavier one (see \cite{Bena:2025usu} for further discussion).

\section{The tower of regular solutions}
\label{sec:TowerSol}

In the previous section, we demonstrated the existence of smooth, asymptotically flat, and rotating topological stars, the regularity of which requires the conditions~\eqref{eq:RangeParam2}. The solutions are given in Section~\ref{sec:TheSol}, and regularity conditions require the spin parameter $a$ to take values in a discrete tower labeled by an integer $k$:
\begin{equation}
     |a| \= \frac{k(r_+-r_-)}{2}\,. \label{eq:RegCondRecap}
\end{equation}
The remaining parameters are determined in terms of the radius of the fifth dimension, $R_\psi$, the quantized KKm charge $N$, the primary integer $k$, and the additional integer $\ell$, as specified in~\eqref{eq:RegQuantization}.

In this section, we analyze the tower of regular solutions labeled by $k$:
\begin{itemize}
    \item \underline{Lowest state at $k=0$ and the static limit:} For $k=0$, we have $a=0$ and the angular momentum vanishes, as shown in~\eqref{eq:ConservedCharges}. 
    \item  \underline{Second state at $k=1$ and the extremal limit:} For $k=1$, one finds $r_+ - r_- = 2 |a|$. In this limit, the solution is extremal and we will show that it is related to a BPS configuration in five-dimensional supergravity.
    \item  \underline{Generic states for $k\geq 2$:} These correspond to generic non-extremal, rotating topological stars.
\end{itemize}

\subsection{Static solutions at $k=0$}
\label{sec:StaticSol}

We consider the static limit $a = 0$ of the solution introduced in Section~\ref{sec:TheSol}. In this limit, the fields simplify, and the solution reads:
\begin{align}
ds_5^2 \= &\frac{ (r - r_+)  \cI_4}{Z^2}\,\left[d\psi - \frac{\chi}{\cI_4}\,dt+ P_0(1+\cos \theta) d\phi \right]^2+ \frac{Z}{\sqrt{(r - r_+)  \cI_4}} \,ds_4^2\,, \nn\\
A \= &  \frac{F_4}{Z} \,dt + \frac{F_2}{Z} \,(d\psi+P_0(1+\cos \theta) d\phi) + \frac{q \,{m^{-}_{1,6}}^2 (r_++r_-)}{(1-q^2){m^{+}_{1,6}}^2}  (1+\cos \theta) \,d\phi\,,
\end{align}
where the four-dimensional spacetime is
\begin{equation}
ds_4^2 \= - \frac{(r-r_-)\sqrt{r-r_+}}{\sqrt{\cI_4}}\,dt^2 + \sqrt{(r-r_+)\cI_4} \left[ \frac{dr^2}{(r-r_+)(r-r_-)} +d\theta^2 +\sin^2 \theta \, d\phi^2\right],
\end{equation}
 the fields are given by
\begin{align}
\cI_4 \= &(r - r_+)^2  \left\{ r - r_- + \frac{ \left[2  m^{-}_{1, 4} + q^2  (r_+ - r_-) \right]  {m_{1, 2}^{(+)}}^2   {m_{1, 6}^{(-)}}^2 }{(1 - q^2)  {m_{1, 6}^{(+)}}^2  m_{2, 6}^{(-)} } \right\}  \nn \\
&+ \frac{3(r_+-r_-) m_{1,2}^{+} {m^{-}_{1,4}}^2 {m^{-}_{1,6}}^2}{(1-q^2)^2 {m^{+}_{1,6}}^2 m^{-}_{2,6}} (r-r_+) + \frac{m^{-}_{1,0}{}^{2} {m^{-}_{1,4}}^3 {m^{-}_{1,6}}^2}{(1-q^2)^3 {m^{+}_{1,6}}^2 m^{-}_{2,6}} \,,\nn\\
Z \= & (r-r_+) \left[r-r_- +\frac{(r_++r_-) (1+q^2) {m^{-}_{1,6}}^2}{(1-q^2) {m^{+}_{1,6}}^2}  \right] + \frac{(r_+-r_-) m^{-}_{1,4} {m^{-}_{1,6}}^2}{(1-q^2)^2 \,{m^{+}_{1,6}}^2}  \,, \\
\chi \= &\frac{q^3 r_- r_+ m^{-}_{1,6}}{{m^{+}_{1,6}}^2 m^{-}_{2,6}} \left\{ (r-r_-) \left[ \left(r-\frac{m^{+}_{1,0}}{2}  \right) \left(2m^{-}_{2,6}+m^-_{1,0} m^{+}_{1,6}  \right) - \frac{3(1+q^2) {m^{-}_{1,0}}^2 m^{-}_{1,6}}{2(1-q^2)} \right] \right. \nn \\
& \hspace{2.5cm} +\left. \frac{(1-q^6) {m^{-}_{1,0}}^3 {m^{-}_{1,6}}^2}{(1-q^2)^3 m^{+}_{1,6}} \right\} \nn \\
F_4 \= & \frac{2q^2 m^{-}_{1,6} \sqrt{r_- r_+ m^{-}_{2,6}}}{\sqrt{1-q^2}\,{m^{+}_{1,6}}^2} \left[r-r_- - \frac{(r_+-r_-) m_{1,2}^{-} m^{-}_{1,6}}{2(1-q^2) m^{-}_{2,6} } \right] \,,\nn \\
F_2 \= & -\frac{2q(1+q^2) m^{-}_{1,6} \sqrt{r_- r_+ m^{-}_{2,6}}}{\sqrt{1-q^2}\,{m^{+}_{1,6}}^2} \left[r-r_+ + \frac{(r_+-r_-) m^{-}_{1,4} m^{-}_{1,6}}{2(1-q^4) m^{-}_{2,6} } \right] \,.\nn 
\end{align}
The conserved charges are
\begin{align}
P_0  &\=  \frac{2 \left(1 -q^6\right) m^{-}_{1,6} \sqrt{ r_+ r_- m^{-}_{2,6}}  }{ (1-q^2)^\frac{3}{2} \, {m^{+}_{1,6}}^2  } ,\qquad P \=  -\frac{q(r_++r_-) {m^{-}_{1,6}}^2 }{(1-q^2){m^{+}_{1,6}}^2} ,\nn\\
Q_0 &\=-\frac{q^3 r_- r_+ m^{-}_{1,6}}{{m^{+}_{1,6}}^2 m^{-}_{2,6}}  \left(2m^{-}_{2,6}+m^-_{1,0} m^{+}_{1,6}  \right)  ,\quad Q  \=- \frac{2q^2 m^{-}_{1,6} \sqrt{r_- r_+ m^{-}_{2,6}}}{\sqrt{1-q^2}\,{m^{+}_{1,6}}^2}  ,   \label{eq:ConservedChargesStatic} \\
M& \=\frac{1}{4} \left[ r_- -r_+ + \frac{ \left(2  m^{-}_{1, 4} + q^2  m^-_{1,0} \right)  {m_{1, 2}^{(+)}}^2   {m_{1, 6}^{(-)}}^2 }{(1 - q^2)  {m_{1, 6}^{(+)}}^2  m_{2, 6}^{(-)} }  \right]\,.
\end{align}
The smooth bolt structure and the spacelike coordinate degeneracy at $r = r_+$ is clearer in the static limit: the degenerating direction is $\psi$ at fixed $\phi$.

When $r_- = 0$, only the magnetic charge $P$ remains. In this case,  $m_{i,j}^{(\pm)} = r_+^i$, and the solution reduces to the static topological star~\cite{Bah:2020ogh,Bah:2020pdz}, with $(r_\text{B},r_\text{S})=\tfrac{r_+}{1-q^2}(1,q^2)$. Our static solution thus generalizes the topological star to a smooth dyonic configuration with an additional parameter, $r_-$. Note that this dyonic solution is also different from the $\cW$-soliton~\cite{Chakraborty:2025ger,Dima:2025tjz}, obtained from a different set of sigma-model transformations.

\subsubsection{Regularity conditions}

In Section \ref{sec:RegCondTS}, we derived the regularity condition at the bolt for the solutions with $a \neq 0$, noting afterward that these conditions remain valid in the limit $a=0$. Here, we will confirm this by starting from scratch with $a=0$ and show that it indeed reproduces the $a=0=k$ conditions of \eqref{eq:RegCond}.

As in the rotating case, there are three main coordinate degeneracies with two intersections: at the poles of the $S^2$ ($\theta = 0, \pi$), at the bolt ($r = r_+$), and at the NUT centers ($r = r_+$ and $\theta = 0, \pi$).

The conditions at the poles of the $S^2$ are identical to those of the rotating geometry: we impose the periodicities given by~\eqref{eq:AngleIdentification} and the quantization of the KKm charge~\eqref{eq:KKmcharge}.

At the bolt, the degenerating direction is $\psi$ at fixed $\phi$. The absence of twisted angles simplifies the analysis: smoothness at the bolt only requires fixing $R_\psi$ in terms of the parameters, without introducing a quantization condition like~\eqref{eq:RegCond}. Defining the local radial distance to the bolt as $\rho^2 = 4 (r - r_+)$, the $(\rho, \psi)$ subspace is:
\begin{equation}
ds_{\rho,\psi}^2 \= d\rho^2 + \rho^2 \,\frac{(1-q^2)^3 \, {m^{+}_{1,6}}^4}{4m^{-}_{2,6} {m^{-}_{1,6}}^4}\,\left[ d\psi +P_0 (1+\cos \theta)d\phi\right]^2\,.
\end{equation}
Thus,  we can impose this degeneracy to correspond to a $\IR^2/\mathbb{Z}_\ell$ space, which, along with the quantization of the KKm charge \eqref{eq:KKmcharge}, requires:
\begin{equation}
\ell\, R_\psi \=  \frac{2  {m^{-}_{1,6}}^2 \sqrt{m^{-}_{2,6}}}{(1-q^2)^\frac{3}{2}\,{m^{+}_{1,6}}^2}, \qquad N R_\psi \= \frac{4 \left(1 -q^6\right) m^{-}_{1,6} \sqrt{ r_+ r_- m^{-}_{2,6}}  }{ (1-q^2)^\frac{3}{2} \, {m^{+}_{1,6}}^2  },
\label{eq:RegCondRpsiStatic}
\end{equation}
which is consistent with the regularity condition~\eqref{eq:RegQuantization} for $a = k = 0$ as expected. The quantization of $r_-$~\eqref{eq:RmQuantization} for $k=a=0$ gives
\begin{equation}
 r_- \= \left\{\sqrt{1+\left[\frac{\ell(1-q^6)}{N q^3} \right]^2} -\frac{\ell(1-q^6)}{N |q|^3}\right\}^2  \frac{r_+}{q^6}\,,
\label{eq:QuantizationRm}
\end{equation}
which directly satisfies the regularity bound~\eqref{eq:RangeParam2}, guaranteeing the absence of CTCs. Moreover, requiring $r_- \leq r_+$ implies:
\begin{equation}
0 \,\leq\, \frac{N}{\ell} \,\leq\, 2\,,
\end{equation}
with $N = 2 \ell$ corresponding to the extremal limit $r_- = r_+$ at $a = 0$. If we seek solutions without an orbifold action ($\ell = 1$), only two cases arise: $N = 1$, yielding $r_- \geq r_+/4$, or $N = 0$, corresponding to the original topological star of~\cite{Bah:2020ogh,Bah:2020pdz} with $r_- = 0$.

\subsubsection{Solitons in the black hole regime} 
\label{sec:BHRegime}

The most general nonextremal black string with electric and magnetic charges in minimal $\mathcal{N}=2$ five-dimensional supergravity is known~\cite{Chow:2014cca}. An important question is whether our static solitons can exist within the same mass and charge range as such black strings.

Unlike the original topological star of~\cite{Bah:2020ogh}, no common parametrization exists between our static solitons and the nonextremal black string, complicating the analysis. We first check whether the solitons satisfy the BPS bound, a necessary, but not sufficient, condition for overlap in the mass/charge range of the black hole, and then numerically test whether a black hole exists with the same mass and charges as our solitons. 

For our charge conventions, the BPS bound of~\cite{Chow:2014cca} translates into:
\begin{equation}
B \equi M^2\- \frac{1}{16} \left[(3P-Q_0)^2+(3Q+P_0)^2 \right] \,\geq\, 0 \,.
\label{eq:BPSBound}
\end{equation}
For our static solutions this simplifies to:
\begin{equation}
B \= \frac{m^{-}_{1,0}{}^{2}}{16} \left[1-\frac{3q^2 {m_{1,2}^{+}}^2{m^{-}_{1,6}}^2}{(1-q^2) m^{-}_{2,6} {m^{+}_{1,6}}^2} \right]\,,
\end{equation}
As expected, $B = 0$ in the extremal limit $r_- = r_+$. Moreover, we find:
\begin{equation}
\begin{split}
&|q| > \frac{1}{2} \qquad \Rightarrow \qquad B < 0\quad \text{for all }r_\pm\,,\\
&|q|< 0.439 \qquad \Rightarrow \qquad B > 0\quad \text{for all }r_\pm\,, \\
&0.439<|q| < \frac{1}{2}\qquad \Rightarrow \qquad B > 0 \quad \text{for }r_-<r_c(q), \quad\text{and } B < 0 \quad  \text{for }r_->r_c(q)\, ,
\end{split}
\label{eq:BPSboundStatic}
\end{equation}
where $r_c(q)$ is a root of a degree-4 polynomial.

However, the BPS bound does not coincide with the black hole extremality bound for generic charge lattices in STU supergravity, so BPS-satisfying solutions can lie outside the black hole regime. Through a numerical scan using the black hole solutions of~\cite{Chow:2014cca}, we noticed that many static solitons coexist with black strings of the same mass and charges, but some BPS-satisfying solitons do not.

Roughly, we find that regular static topological stars with $|q| < \frac{1}{2}$ and
\begin{equation}
    0 \,\leq\, \frac{r_-}{r_+} \,\lesssim\, \begin{cases}
        e^{-4 |q|}\, ,\quad |q|<0.4\,,\\
        1-2|q| \,,\quad 0.4<|q|< \frac{1}{2},
    \end{cases}
\end{equation}
have the same mass and charges as a non-extremal static black string in five-dimensional supergravity. As in the case of the topological star of~\cite{Bah:2020ogh,Bah:2020pdz}, there exists a parameter range in which the solitons share the same conserved quantities of a non-extremal black hole and can thus represent one of its microstates. Outside this range, the solitons correspond to smooth horizonless geometries without a black hole counterpart.

\subsection{Extremal solutions at $k=1$}
\label{sec:ExtremalLimit}

We now consider the $k=1$ solution, which yields $2|a| = r_+ - r_-$. As argued in~\eqref{eq:ExtremalLimit}, this corresponds to the extremal limit where the three-dimensional base becomes flat $\IR^3$. We therefore expect the topological star to become BPS in this limit.

For simplicity, we take $2a = r_- - r_+$.\footnote{Taking $2a = r_+ - r_-$ would lead to the same analysis upon exchanging the North and South pole coordinates~\eqref{eq:NorthSouthCoor}.} Depending on the sign of $r_+ + r_-$, we must express the solution using either the North-pole coordinates~\eqref{eq:NorthSouthCoor} (for $r_+ + r_- > 0$) or the South-pole coordinates (for $r_+ + r_- < 0$). In both cases, the metric and fields simplify to
\begin{align}
    ds^2 &= \frac{H_1}{H_2^2} \left[d\psi+\frac{Q_0}{r_X H_1} dt+ P_0(\epsilon +\cos \theta_X)d\phi \right]^2 - \frac{H_2}{H_1} dt^2+ H_2 \,\left[ dr_X^2 +r_X^2 \left(d\theta_X^2 + \sin^2 \theta_X \,d\phi^2 \right)\right], \nn\\
    A &= -\frac{Q}{r_X H_2} dt +\frac{(1+q^2) Q}{q r_X H_2} \left[ d\psi+ P_0(\epsilon+\cos \theta_X)d\phi \right] -P (1+\cos \theta_X)d\phi, \label{eq:BPSlimit}
\end{align}
where $X = N$ and $\epsilon = 1$ (resp. $S$ and $\epsilon = -1$) for $r_+ + r_- > 0$ (resp. $<0$). The functions and charges are
\begin{align}
    P_0 &= \frac{(1+q^2+q^4)^\frac{3}{2}(r_++r_-)m^{-}_{2,6}}{{m^{+}_{1,6}}^2},\quad P=-\frac{\epsilon q}{\sqrt{1+q^2+q^4}} P_0,\quad Q \=- \frac{P^2}{P_0},\quad Q_0 \= \frac{P^3}{P_0}, \nn\\
    H_1 &\= 1+ \frac{(1+q^2)^3 P_0}{(1+q^2+q^4)^\frac{3}{2} \,r_X},\qquad H_2 \= 1+ \frac{(1+q^2)P_0}{\sqrt{1+q^2+q^4}\,\,r_X}. \label{eq:ConservedchargesBPS}
\end{align}
This solution is regular everywhere, with the smooth structure of Taub-NUT space and a regular center at $r_X = 0$ with NUT charge $P_0$.

The solution is static and saturates the BPS bound~\eqref{eq:BPSBound}, with
\begin{equation}
    M \= \frac{(1+q^2)^3(r_++r_-)m^{-}_{2,6}}{4m^{+}_{1,6}}.
\end{equation}
The mass is positive for any solution with $r_+ + r_- \geq 0$, or with $r_+ + r_- < 0$ and $m^{-}_{2,6} < 0$, that is $r_- < - |q|^{-3} r_+$, and we retrieve the validity bound~\eqref{eq:RangeParam2}.

The solution can be recast in the Bena-Warner ansatz~\cite{Bena:2005va,Bena:2007kg} by redefining the fibers:
\begin{equation}
\begin{split}
    ds^2 =&  \,-\frac{1}{Z^{2}} \left\{dt + \mu \left[d\psi+\frac{q(3+2q^2+3q^4)}{(1+q^2)^3} dt + P_0(\epsilon+\cos \theta_X)d\phi\right] \right\}^2 +Z ds_4^2\,, \\
   A =& \,\frac{1}{Z} \, \left\{ dt + \mu \left[d\psi+\frac{q(3+2q^2+3q^4)}{(1+q^2)^3} dt + P_0(\epsilon+\cos \theta_X)d\phi\right]\right\}\\
   &\,-\frac{K}{Z_0}  \left[d\psi + \frac{q(3+2q^2+3q^4)}{(1+q^2)^3} dt + P_0(\epsilon+\cos \theta_X)d\phi\right]  -P(1+\cos \theta_X)d\phi\,, 
\end{split}
\end{equation}
where $ds_4^2=\frac{1}{Z_0} (d\psi +\frac{q(3+2q^2+3q^4)}{(1+q^2)^3} dt + P_0(\epsilon+\cos \theta_X)d\phi)^2 + Z_0\,ds(\IR^3)^2$ is a hyper-Kähler base and the fields are given in terms of four harmonic functions $(Z_0,K,L,M)$:
\begin{equation}
    Z \= L+\frac{K^2}{V},\qquad \mu \= \frac{M}{2} + \frac{3K L}{2V}+\frac{K^3}{V^2},
\end{equation}
with
\begin{align}
    Z_0 &\= \frac{(1-q^2)^2\sqrt{1+q^2+q^4}}{(1+q^2)^3}+\frac{P_0}{r_X},\qquad K \= \frac{q(3+2q^2+3q^4)}{(1+q^2)^3} - \frac{P}{r_X},\\
    L&\= \frac{(1-q^2)^2\sqrt{1+q^2+q^4}}{(1+q^2)^3}-\frac{Q}{r_X},\qquad M \= -\frac{q(3+2q^2+3q^4)}{(1+q^2)^3}+\frac{Q_0}{r_X}. \nn
\end{align}
The harmonic functions correspond to a single center, with electric charges determined by the magnetic charges as in~\eqref{eq:ConservedchargesBPS}, characteristic of BPS Gibbons-Hawking centers preserving 16 supercharges.

Thus, the coordinates that make the BPS nature of the solution manifest have required shifting the static frame~\eqref{eq:BPSlimit} so that the angle forming the hyper-Kähler base is $\psi+\frac{q(3+2q^2+3q^4)}{(1+q^2)^3} \,t$. This corresponds to a \emph{spectral flow}, as extensively studied in the context of BPS microstate geometries~\cite{Bena:2008wt}. In this BPS frame, the solution is not asymptotically static but rotates along the fifth dimension, as for the black ring in Taub-NUT. Returning to the asymptotically static frame gives the form~\eqref{eq:BPSlimit}.

We have thus shown that the $k=1$ solutions are static and correspond to the spectral flow of a BPS configurations of five-dimensional supergravity sourced by a single Gibbons-Hawking center.

\subsection{Generic rotating solutions at $k\geq 2$}
\label{sec:RotatingSol}

Generic solutions with $k \geq 2$ are determined by the metric and fields given in Section~\ref{sec:TheSol}. Subject to the regularity conditions and compatibility with asymptotic flatness, the main parameters $(r_+, r_-, a)$ are specified by the transformation parameter $q$, the radius of the extra dimension $R_\psi$, and three integers $(k, \ell, N)$ through~\eqref{eq:RmQuantization} and~\eqref{eq:RegQuantization}. In addition, the parameters must satisfy the condition~\eqref{eq:RangeParam2}, ensuring the solution is free of CTCs, has positive mass, and does not have regions of signature change.

The solutions are rotating and carry all possible charges allowed in five-dimensional supergravity for geometries asymptotic to S$^1 \times \mathbb{R}^{1,3}$: Kaluza-Klein monopole (KKm) and momentum (P) charges along the S$^1$, as well as magnetic and electric charges under the $U(1)$ gauge field~\eqref{eq:ConservedCharges}.

A key question is whether a nonextremal rotating black string exists in the same regime of mass, charges, and spin as the topological star, as it happens in the absence of spin (Section~\ref{sec:BHRegime}). To investigate this, we consider the general STU black hole of~\cite{Chow:2014cca} and its embedding and consistent truncation to minimal five-dimensional supergravity. We follow the same approach as for the static solutions: we first check whether the rotating topological star satisfies the BPS bound~\eqref{eq:BPSBound}, and then numerically scan for black hole solutions with matching mass, charges, and spin. For the rotating topological star, the BPS bound simplifies drastically,
\begin{equation}
B \= \frac{m^{-}_{1,0}{}^{2}-4a^2}{16} \left[1-\frac{3q^2 {m_{1,2}^{+}}^2({m^{-}_{1,6}}^2-4a^2 q^6)}{(1-q^2) m^{-}_{2,6} {m^{+}_{1,6}}^2} \right]\,. 
\end{equation}
and the parameter $a$ is fixed in terms of $r_\pm$ by~\eqref{eq:RegCondRecap}.\footnote{Although $r_\pm$ are in principle fixed by~\eqref{eq:RegQuantization}, we can treat them as free parameters within the bound \eqref{eq:RangeParam2} since they can be varied independently by adjusting $(N, \ell)$.}

We have verified that solutions in the validity range~\eqref{eq:RangeParam2} always satisfy the BPS bound $B>0$. However, our numerical search for black hole solutions with the same mass, charges, and spin as the regular topological stars was unsuccessful. While not definitive, this suggests that our rotating topological stars do not exist in the same mass-charge-spin regime as nonextremal rotating black holes of STU supergravity: the \emph{rotating topological star, unlike its static limit, lies outside the black hole extremality bound}.

This feature appears related to the fact that the rotating topological stars are generated by applying sigma-model transformations to a Kerr-Taub-bolt geometry, which for $k \geq 2$ is itself outside the extremality bound of the black hole with the same mass and KKm charge ($|n|>m$, see Appendix~\ref{sec:Kerr-Taub-bolt}). While it is known that applying a sigma-model transformation to a black hole seed does not move the solution in or out of the extremality bound~\cite{Chow:2014cca}, it is possible that this also holds when applied to a smooth horizonless geometry.

Thus, the rotating topological stars constructed here join previously known smooth solutions, such as JMaRT~\cite{Jejjala:2005yu} and its floating JMaRT extensions~\cite{Bossard:2014yta,Bena:2015drs,Bena:2016dbw,Bossard:2017vii}, in lying outside the black hole bound. However, unlike those solutions, this is not due to over-rotation but rather to an excess of charges (the Kerr-Taub-bolt seed requires $|n| > m$). Moreover, unlike those cases, we know how to resolve this: we must start from a seed solution that lies within the black hole bound, which will be the subject of future studies.

\section{Rotating topological stars close to the boosted Kerr black string}
\label{sec:TSclosetoKerr}

Despite the fact that rotating topological stars cannot coexist with black holes, we can construct solutions that have \emph{almost} the same mass, charges, and spin as a non-extremal black hole. From a phenomenological perspective, these configurations can therefore be meaningfully compared.

As an example of such configurations, we analyze the interesting near-vacuum limit of topological stars, already mentioned in~\eqref{eq:VacuumLimit}. This limit is obtained by taking $r_-$ close to $-|q|^{-3} r_+$ while satisfying the regularity bound~\eqref{eq:RangeParam2},\footnote{Note that $r_-$ is a priori fixed in terms of $(q, N, \ell, r_+)$ by~\eqref{eq:RmQuantization}. However, one can treat $r_-$ as a free parameter by suitably adjusting $q$, $N$, and $\ell$. For instance, we can make $r_-$ arbitrarily close to $-|q|^{-3} r_+$ by choosing $q$, $N$, and $\ell$ such that:
\begin{equation}
-2 \ell + k N - \frac{N \left( k^2 (1 + q^6) - 2 |q|^3 \right)}{
(1 - |q|^3) \sqrt{ k^2 (1 + |q|^3)^2 - 4 |q|^3 }} = \mathcal{O}(\epsilon).
\end{equation}}
\begin{equation}
    r_- \= -\frac{1+\epsilon}{|q|^3}\,\,r_+\,,\qquad \epsilon \ll 1\,,
    \label{eq:VacLim}
\end{equation}
which implies $m_{2,6}^- \sim -2 r_+^2 \epsilon$. Without loss of generality, we assume $0 \leq q < 1$ for simplicity. The leading-order contributions to the mass, charges, and spin of the topological star are:
\begin{align}
P_0 &\= \cO(\sqrt{\epsilon})\,,\qquad Q \= \cO(\sqrt{\epsilon}), \qquad J \= - \frac{k (k^2 - 1)^{3/2} (1 + q^3)^4 \,r_+^2}{4 \sqrt{2} q^6 (1 - q^2)^{3/2} (1 - q^3) \sqrt{\epsilon}},\nn \\
Q_0 &\= \frac{(k^2 - 1)(1 + q^3)^2 r_+}{2 q^3 (1 - q^3)\, \epsilon} 
\left[ 1 + \frac{(3 - 6 q^3 - q^6)\epsilon}{2 (1 - q^3)(1 + q^3)} \right],\qquad P \= \frac{(1-k^2)(1 + q^3)^2 r_+}{(1 - q^3)(1 - q^2) q^2},\nn \\
M &= \frac{(k^2 - 1)(1 + q^3)^2 \,r_+}{8 q^3 (1 - q^3) \epsilon} 
\left\{ 1 + \frac{9 + q^2 \left[ 12 - q \left(6 - 12 q - 5 q^3 \right) \right]}{2 (1 - q^6)} \epsilon \right\}. \label{eq:ConservedChargesVacLim}
\end{align}
The regularity condition~\eqref{eq:RegCond}, associated with the quantization of the KKm charge $P_0$, yields at leading order in $\epsilon$:
\begin{equation}
R_\psi \= \frac{
2 \sqrt{2} (1 + q^3)^2 \sqrt{ (k^2-1)\left[ k^2 (1 + q^3)^2 -4 q^3\right]}
}{
N \, q^3 (1 - q^2)^{3/2} (q^3-1)
} \,\sqrt{\epsilon}\,r_+\,,
\end{equation}
which is well-defined for $N\leq -1$. Moreover, we observe that $r_+$ (and thus the ADM mass of the topological star) decouples from the radius of the extra dimension and can be made arbitrarily large in the limits of large $|N|$ and small $\epsilon$. This decoupling is not possible for static solutions, including the original topological star of~\cite{Bah:2020ogh}, where such behavior can only be achieved by introducing a conical defect at the bolt.

\subsection{Comparison to the boosted Kerr black hole}

If we focus only on the dominant conserved charges, namely, the ADM mass $M$, the momentum charge $P$, and the angular momentum $J$, the topological star has the same leading-order charges as a vacuum black string: the boosted Kerr black string in five dimensions. This black string is obtained by trivially embedding the Kerr black hole in five dimensions and applying a Lorentz boost along the fifth dimension. The metric is given by:
\begin{align}
    ds_\text{BH}^2 = &\left[1+\frac{2m q_1^2\,r}{(1-q_1^2)(\Sigma+2m r)} \right] \left[ d\psi+ \frac{2 m q_1\,r}{(1-q_1^2) \Sigma+2m r} \left(dt+\omega_t\right)+\omega_\psi\right]^2 \\
    &- \left[1-\frac{2m \,r}{(1-q_1^2)\Sigma+2m r} \right]\left(dt+\omega_t\right)^2 + \left(1+\frac{2m\,r}{\Sigma} \right) \left(\frac{\Sigma\,dr^2}{\Delta}+\Sigma\,d\theta^2+\Delta \sin^2\theta \,d\phi^2 \right),\nn
\end{align}
with
\begin{equation}
    \omega_t = \frac{2 a m r \sin^2 \theta}{\sqrt{1-q_1^2} \Sigma}d\phi ,\quad \omega_\psi = -\frac{2 a m q_1 r \sin^2 \theta}{\sqrt{1-q_1^2} \Sigma}d\phi, \quad \Delta = r^2-2m r+a^2,\quad \Sigma= \Delta-a^2\sin^2\theta,
    \label{eq:QuantitiesBoostedKerr}
\end{equation}
and where $q_1 = \tanh \delta_1$ is the boost parameter, with $q_1 = 0$ yielding the unboosted Kerr metric trivially fibered over S$^1$, and the range $|q_1| < 1$.

The conserved charges of this black string are:
\begin{equation}
    M^\text{BH} \= \frac{m}{4}\left( 3+\frac{1+q_1^2}{1-q_1^2}\right),\qquad J^\text{BH}=\frac{am}{\sqrt{1-q_1^2}},\qquad Q_0^\text{BH} \= \frac{2m q_1}{1-q_1^2}.
\end{equation}
The BPS (and extremal) limit is reached by taking $q_1 \to 1$ (infinite boost) while sending $m, a \to 0$ such that $M^\text{BH}$ and $Q_0^\text{BH}$ remain finite with $M^\text{BH}\to 4 Q_0^\text{BH}$, and $J^\text{BH} \to 0$.

Interestingly, the boosted Kerr black hole reproduces the leading-order conserved charges of the topological star~\eqref{eq:ConservedChargesVacLim} if we take only a partial extremal limit: $q_1 = 1 - \cO(\epsilon)$ but $m$ and $a$ kept finite. This corresponds to approaching the BPS limit for the momentum charge, where $M^\text{BH} \sim 4 Q_0^\text{BH}$, while keeping a large angular momentum $J^\text{BH}$ consistent with~\eqref{eq:ConservedChargesVacLim}. The momentum charge, mass and spin match precisely under the identification:
\begin{equation}
    q_1 \= 1-\frac{1+q^2}{1-q^2}\epsilon,\quad m \= \frac{(k^2-1)(1+q^2)(1+q^3)^2 r_+}{2 q^3(1-q^3)(1-q^2)}, \quad a=-\frac{k(1-q^3)}{\sqrt{(k^2-1)(1+q^2)^3}} \,m\,.
\end{equation}
The black string remains within its extremality bound if $|a|\leq m$, which requires
\begin{equation}
    k^2 \,\geq\, \frac{(1 + q^2)^3}{q^2 [3 + q (2 + 3 q)]},
    \label{eq:BoostedKerrCond}
\end{equation}
a condition automatically satisfied for $k \geq 2$ when $q \geq 0.288$, but which restricts the allowed $k$ for smaller $q$.

Therefore, near the vacuum limit~\eqref{eq:VacLim}, topological stars exhibit the same mass, momentum, and angular momentum to those of a highly boosted Kerr black hole, provided the above condition holds, while all the other charges are negligible. Beyond the conserved charges, it is natural to ask to what extent the gravitational moments of the topological stars differ from those of the boosted Kerr. The nonvanishing multipole moments of a boosted Kerr are given by~\cite{Bena:2020uup}: 
\begin{equation}
    M_{2n}^\text{B-Kerr}= M^\text{BH} (-a^2)^n,\qquad S_{2n+1}^\text{B-Kerr}= J^\text{BH} (-a^2)^n,
\end{equation}
while those of the topological stars are given in~\eqref{eq:MultipolesTS}.

As discussed in Section~\ref{sec:Multipoles}, the mass quadrupole of topological stars can be positive. Yet, within the range~\eqref{eq:VacLim}, it is negative, and comparable to the mass quadrupole of a boosted Kerr black hole with the same mass, momentum, and angular momentum. At leading order in $\epsilon \to 0$, one finds
\begin{equation}
    \frac{M_2}{M_2^\text{B-Kerr}} = \frac{(1 + q + q^2)^2}{(1 + q)^2 (1 + q^2)}\,,
\end{equation}
which interpolates between $1$ for $q \to 0$ and $9/8 \approx 1.12$ for $q \to 1$. However, unlike the boosted Kerr, the topological star exhibits nonzero odd mass multipoles and even spin multipoles, which are large in the $\epsilon \to 0$ limit. Consequently, the multipole structure of topological stars remains significantly different from that of the boosted Kerr geometry.

Nonetheless, from a phenomenological perspective, the relevant quantities are the \emph{dimensionless} moments $M_j/M^{j+1}\sim \epsilon^j\ll1$ and $S_j/M^{j+1}\sim \epsilon^{j+1/2}$, all of which remain small in the $\epsilon\ll1$ limit. This includes the dimensionless
spin $S_1/M^2\sim \epsilon^{3/2}$. Hence, in this regime, rotating topological stars are slowly spinning, in contrast to the JMaRT solution, which carries large angular momentum above the black hole bound. This supports earlier observations that our topological star exceeds the black hole bound because of its charges rather than its angular momentum.
Overall, while the difference in the multipolar structure relative to that of a boosted Kerr remains sizable and potentially measurable, in this limit the absolute magnitude of the moments is small and can hardly leave an observational imprint.

\subsection{Comparison to the Cveti\v{c}-Youm black hole}

While the previous topological stars match the mass, momentum, and angular momentum of a boosted Kerr black hole, subleading effects differentiate the two solutions: notably, the topological star possesses a magnetic charge $P$ of order $\cO(1)$ (see Eq.~\eqref{eq:ConservedChargesVacLim}) which can be taken into account by refining the comparison to the Cveti\v{c}-Youm black hole~\cite{Cveti__1996,Chong:2004na}. The embedding and truncation of the Cveti\v{c}-Youm black hole in five-dimensional supergravity yield the following solution
\begin{align}
ds_\text{BH}^2 &\= \frac{\cI_4}{Z^2}\,\left[d\psi + \frac{\chi}{\cI_4}\,(dt+\omega_t )+ \omega_\psi \right]^2 \nn\\
& + \frac{Z}{\sqrt{\cI_4}} \left[- \frac{1}{\sqrt{\cI_4}}\,(dt+\omega_t )^2+\sqrt{\cI_4}  \left(\frac{\Sigma\,dr^2}{\Delta}+\Sigma\,d\theta^2+\Delta \sin^2\theta \,d\phi^2 \right) \right], \\
A &=   \frac{2 a m q_2 \left( 1 + q_1 q_2 \right) \cos\theta }{\sqrt{1 - q_1^2} \left(1 - q_2^2\right)^\frac{3}{2}\Sigma Z} \,\left[(dt+\omega_t ) + \frac{q_1+q_2}{1+q_1 q_2} \,(d\psi+\omega_\psi )\right] -P^\text{BH}  \left(1 + \frac{a^2 \sin^2\theta}{\Sigma}   \right)d\phi,\nn
\end{align}
with $\Sigma$ and $\Delta$ as in~\eqref{eq:QuantitiesBoostedKerr} and:
\begin{align}
   Z &\= 1 + \frac{2m \left( r (1 - q_2^4)+2m q_2^4  \right)}{\Sigma \left(1 - q_2^2 \right)^2},\nn\\ 
   \cI_4 &\= 1 + \frac{4 M^{\text{BH}} \,r}{\Sigma}- 4 m^2\frac{ q_2^3 \Sigma \left[ 2 q_1 + q_2^3 + q_1^2 q_2 (3 - 2 q_2^2)\right]-\left[ r + (r-2m) q_1 q_2^3 \right]^2}{(1 - q_1^2)(1 - q_2^2)^3\,\Sigma^2 },\\
   \chi &\= \text{\small $ \frac{ Q_0^{\text{BH}}}{\Sigma} \left\{ r - \frac{ 2m q_2^3 \left[ 1 + q_1 (q_1 + 3 q_2 - q_2^3) \right]}{q_1 (1 - q_2^2)^3}\right\}+\frac{4 m^2 \left[ r(q_1 + q_2^3)-2m q_2^3  \right] \left[  r(1 + q_1 q_2^3)-2m q_1 q_2^3 \right]}{\Sigma^2 (1 - q_1^2)(1 - q_2^2)^3}, $} \nn\\
   \omega_t &\= \frac{2 J^{\text{BH}} \sin^2\theta}{\Sigma} \left( r - \frac{2m q_1 q_2^3}{1 + q_1 q_2^3} \right) d\phi,\quad \omega_\psi \= -\frac{2 a m \sin^2\theta \left[ r \left( q_1 +q_2^3 \right) -2m  q_2^3\right]}{\Sigma \sqrt{1 - q_1^2} \left(1 - q_2^2\right)^{3/2}}d\phi,\nn
\end{align}
with conserved charges:
\begin{equation}
\begin{split}
    P^\text{BH}&\= \frac{2m q_2}{1-q_2^2},\qquad Q_0^\text{BH}\= \frac{2m q_1}{1-q_1^2},\\
    M^\text{BH}&\= \frac{m}{4}\left(\frac{1+q_1^2}{1-q_1^2}+3 \frac{1+q_2^2}{1-q_2^2}\right), \qquad J^\text{BH}\= \frac{am(1+q_1 q_2^3)}{\sqrt{1 - q_1^2} \left(1 - q_2^2\right)^{3/2}}.
\end{split}
\end{equation}
Compared to the boosted Kerr black hole, this solution introduces an extra transformation parameter $-1 < q_2=\tanh \delta_2 < 1$ controlling the magnetic charge.

The black hole reproduces all leading charges of the topological star~\eqref{eq:ConservedChargesVacLim} under the identification:
\begin{align}
    q_1&\=1-\epsilon,\quad q_2 \=-q,\quad m \= \frac{(k^2-1) (1 + q^3)^2 r_+}{2 q^3 (1 - q^3)}\left( 1 + \frac{2 - 3 q^3 - q^6}{1 - q^6} \epsilon \right), \nn\\
    a &\= -\frac{k \sqrt{k^2 - 1} \,(1 + q^3)^2 \,r_+}{2 (1 - q^3) q^3}\,.
\end{align}
As before, the black hole lies within the physical regime if $|a|\leq m$, which requires $k$ to scale as $k \sim \epsilon^{-1/2}$ with $r_+ \sim \epsilon$ in the small $\epsilon$ limit. This gives:
\begin{equation}
    k \= \frac{k_0}{\sqrt{\epsilon}},\qquad r_+ \= \epsilon\,r_0,
\end{equation}
with $2k_0^2 \geq \frac{1 - q^6}{2 - 3 q^3 - q^6}$ and $q < 0.825$. This implies $m \sim |a|$, i.e., the black hole is near extremality. Thus, by taking into account the subleading magnetic charge $P$, the topological star is best compared to a near-extremal Cveti\v{c}-Youm black hole. 

If one wants to go one step further and include the remaining infinitesimal charges $P_0 = \cO(\sqrt{\epsilon})$ and $Q = \cO(\sqrt{\epsilon})$, this would require comparison to the most general black string in five-dimensional minimal supergravity as constructed in~\cite{Chow:2014cca}. However, as previously discussed, our mapping of the topological star parameters to the black string was unsuccessful, indicating that the topological star lies just outside the extremality bound of such black holes for the same charges and spin.\\

In conclusion, we have demonstrated that topological stars close to the vacuum limit~\eqref{eq:VacLim} can meaningfully be compared to a highly boosted Kerr black hole when ignoring subleading charges, and to a near-extremal Cveti\v{c}-Youm black hole when incorporating the small magnetic charge. This motivates future studies of their gravitational signatures to explore potential observational distinctions between these horizonless geometries and the classical black holes.

\section{Gravitational signatures and dynamical properties} \label{sec:GravSign}

Despite the apparent complexity of the rotating topological star~\eqref{eq:RTSMetric&Field}, we show in this section that it possesses all the key properties needed for a tractable analysis of its dynamics under linear perturbations. In particular, we demonstrate that both the geodesic equations for test particles and the Klein-Gordon equation for a minimally coupled scalar field admit separation of variables into integrable radial and angular components. This ensures that studies of photon scattering, photon rings, imaging simulations, classical stability under scalar perturbations, scalar quasinormal modes, and scalar ringdown signals are feasible, and we defer their detailed investigation to future work.

\subsection{Separability of geodesic equations}
The separability of the geodesic equation is most naturally analyzed within the Hamilton-Jacobi formalism.
We begin with the Lagrangian for geodesic motion,
\begin{equation}
\mathcal{L}(x^c,\dot{x}^c) = \frac{1}{2}\, g_{ab}(x^c)\,\dot{x}^a \dot{x}^b\,,
\end{equation}
from which the canonical momenta are obtained as
\begin{equation}
p_a = \frac{\partial \mathcal{L}}{\partial \dot{x}^a} = g_{ab}\,\dot{x}^b\,,
\qquad \Rightarrow \qquad
\dot{x}^a = g^{ab} p_b\,.
\end{equation}
Here, a dot denotes differentiation with respect to the affine parameter $\lambda$. Along any geodesic the Lagrangian takes a fixed value,
\begin{equation}
\mathcal{L}(x^a,\dot{x}^a) = \frac{\sigma}{2}\,,
\end{equation}
where $\sigma =-1,0,1$ for timelike, null, and spacelike geodesics, respectively.
The Hamiltonian associated with $\mathcal{L}$ is
\begin{equation}
\mathcal{H}(x^c,p_c) = \frac{1}{2}\, g^{ab}(x^c)\, p_a p_b\,,
\end{equation}
and the geodesic equations follow from Hamilton’s equations,
\begin{equation}
\dot{x}^a = \frac{\partial \mathcal{H}}{\partial p_a}\,,
\qquad
\dot{p}_a = -\frac{\partial \mathcal{H}}{\partial x^a}\,.
\end{equation}
In the Hamilton-Jacobi formalism, one introduces Hamilton’s principal function $S(x^a,\lambda)$, which depends on the coordinates and the affine parameter. The Hamilton-Jacobi equation then reads
\begin{equation}
\mathcal{H}\left(x^a, \frac{\partial S}{\partial x^a}\right) + \frac{\partial S}{\partial \lambda} = 0\,.
\end{equation}
On a solution of this equation, the canonical momenta are given by the gradients of $S$:
\begin{equation}
p_a = \frac{\partial S}{\partial x^a}
\qquad \Rightarrow \qquad
\dot{x}^a = g^{ab}\frac{\partial S}{\partial x^b}\,.
\end{equation}

Since $\partial/\partial t$, $\partial/\partial \phi$, and $\partial/\partial \psi$ are Killing vector fields, we have
\begin{equation}
\begin{aligned}
\mathcal{H}=\frac{\sigma}{2}\,,\quad p_t=-E\,,\quad p_{\phi}=L_{\phi}\quad\text{and}\quad p_{\psi}=L_{\psi}\,,
\end{aligned}
\end{equation}
with $E$, $L_{\phi}$ and $L_{\psi}$ being constants. It follows that
\begin{equation}
S(x^a,\lambda)=-\frac{\sigma}{2}\lambda-E\,t+L_{\phi}\,\phi+L_{\psi}\,\psi+\hat{S}(r,\theta)\,.
\end{equation}

We assume a separable ansatz for $\hat{S}(r,\theta)$,
\begin{equation}
\hat{S}(r,\theta)=S_r(r)+S_x(\cos\theta),
\end{equation}
and introduce the compact coordinate $x=\cos\theta$. After some algebra, one finds
\begin{equation}
{S_x^\prime(x)}^2=\frac{\Xi_x(x)}{1-x^2}\quad\text{and}\quad{S_r^\prime(r)}^2=\frac{\Xi_r(r)}{\Delta(r)}
\end{equation}
with
\begin{align}
\Xi_x(x)\= &\left(3L_{\psi}^2-\sigma\right)\Delta Z_x(x)-\frac{\left[L_{\phi }-L_{\psi}\,\omega _{\psi \infty}(x)\right]^2}{1-x^2}-(1-x^2)a^2E^2
\nn \\
&-(L_{\psi}^2+E^2)\,\Delta I^{(4)}(x)+2 L_{\psi}E \,\Delta \chi^{(2)}(x)+L_{\psi}(L_{\phi}-L_{\psi}P_0)P_0+\mathcal{C} \,, \nn\\
\Xi_r(r) \= &\sigma\,[Z_r(r)+Z_x(1)]-\frac{a^2(L_{\phi}-L_{\psi}P_0)^2-I(r,0)[E-L_{\psi}\chi(r,0)]^2}{\Delta(r)}
\\
&-\frac{L_{\psi}^2[Z_r(r)+Z_x(1)]^3}{I(r,0)}-(L_{\phi}-L_{\psi}P_0)\left[E\, \partial^2_\theta\omega _t(r,0)-L_{\psi }\,\partial^2_\theta\omega_{\psi}(r,0)\right]-\mathcal{C}, \nn
\end{align}
where $\mathcal{C}$ plays the role of a Carter-like separation constant, and the auxiliary quantities are defined from the fields~\eqref{eq:FieldsRTS1}:
\begin{subequations}
\begin{equation}
\begin{aligned}
&Z_r(r)=\Delta(r)-\frac{(1+q^2) m_{1,0}^+ \left(4 a^2
   q^6-m_{1,6}^-{}^2\right)}{(1-q^2)
   m_{1,6}^+{}^2}r-\frac{q^4
   m_{1,0}^+{}^2 \left(4 a^2
   q^6-m_{1,6}^-{}^2\right)}{\left(1-q^2\right)^2
   m_{1,6}^+{}^2}\,,
\\
&Z_x(x)=a^2 (1-x^2)+\frac{2 a n
   (1+q^2)\left(4 a^2 q^6-m^-_{1,6}{}^2\right)}{\left(1-q^2\right) m^+_{1,6}{}^2}x+\frac{2 n^2 \left(1+q^4\right) \left(4 a^2
   q^6-m^-_{1,6}{}^2\right)}{\left(1-q^2\right)^2 m^+_{1,6}{}^2}\,,
\\
&I(r,\theta)=Z(r,\theta) F_1(r,\theta)-F_2(r,\theta)^2\,, \qquad \chi(r,\theta)=\frac{F_2(r,\theta) F_4(r,\theta)-Z(r,\theta) F_3(r,\theta)}{I(r,\theta)}\,,
\end{aligned}
\end{equation}
and
\begin{equation}
\begin{aligned}
&\omega _{\psi \infty}(x)=\lim_{r\to+\infty}\left.\omega_\psi(r,\theta)\right|_{\theta=\arccos x}\,,\qquad \Delta Z_x(x)=Z_x(1)-Z_x(x),
\\
&\Delta I^{(4)}(x)=I^{(4)}(1)-I^{(4)}(x)\,,\qquad I^{(4)}(x)=\left.\frac{1}{2}\lim_{r\to+\infty}r^4\left[\frac{\partial^2}{\partial r^2}\left(\frac{I}{r^4}\right)+\frac{2}{r}\frac{\partial}{\partial r}\left(\frac{I}{r^4}\right)\right]\right|_{\theta=\arccos x},
\\
&\Delta \chi^{(2)}(x)=\chi^{(2)}(1)-\chi^{(2)}(x)\,,\qquad \chi^{(2)}(x)=\left.-\lim_{r\to+\infty}r^2\frac{\partial\left(r\chi\right)}{\partial r}\right|_{\theta=\arccos x}.
\end{aligned}
\end{equation}
\end{subequations}

The geodesic equations follow directly by relating the momenta to the velocities:
\begin{equation}
\dot{r}^2=\frac{\Delta(r)\Xi_r(r)}{Z(r,\theta)^2}\quad\text{and}\quad \dot{\theta}^2=\frac{\Xi_x(\cos \theta)}{Z(r,\theta)^2}\,.
\end{equation}

The existence of a separation constant $\mathcal{C}$ implies the existence of a Killing tensor with components $K_{ab}$, analogous to the Carter-Penrose-Walker tensor for the Kerr black hole~\cite{carter_1968,walker_penrose_1970}. In principle, one can read off the components of this tensor by requiring
\begin{equation}
    K^{ab} p_a p_b = \mathcal{C}\,.
\end{equation}
This tensor ensures that the massive scalar wave equation is separable~\cite{Carter1968CMP}, as we demonstrate explicitly in the next section.  
A natural question is whether an associated Killing-Yano tensor exists.  
If such a tensor is present, its existence would guarantee that Proca test fields (including the massless limit) are also separable~\cite{Frolov:2018ezx}.  
While it seems plausible that a Killing-Yano tensor exists for this spacetime, we leave a detailed investigation of this question for future work.

\subsection{Separability of scalar perturbations}
Consider the five-dimensional Klein-Gordon equation for a massive test scalar field~$\Phi$ with mass~$\mu$:
\begin{equation}
\Box \Phi =\mu^2\Phi\,.
\end{equation}
We introduce a \emph{separable} ansatz for $\Phi$:
\begin{equation}
\Phi(t,r,\theta,\phi,\psi)=X(\cos \theta)R(r)e^{-i \omega t+i m_{\psi}\psi+im_{\phi}\phi}\,.
\end{equation}
A straightforward, though algebraically tedious, calculation shows that the functions $X(x)$ and $R(r)$ satisfy the following ordinary differential equations:
\begin{subequations}
\begin{equation}
\begin{aligned}
&\partial_x[(1-x^2)\partial_xX(x)]+V_x(x)X(x)=0
\\
&\partial_r[\Delta(r)\partial_rR(r)]+V_r(r)R(r)=0
\end{aligned}
\end{equation}
with
\begin{multline}
V_x(x)=(3m_{\psi}^2+\mu^2)\Delta Z_x(x)-\frac{\left[m_\phi-m_\psi\, \omega_{\psi\infty}(x)\right]^2}{1-x^2}-(1-x^2)a^2\omega^2
\\
-\left(m_{\psi}^2+\omega^2\right)\Delta I^{(4)}(x)+2 m_{\psi}\omega \Delta \chi^{(2)}(x)+m_{\psi}(m_{\phi}-2m_{\psi}P_0)P_0+\Lambda
\end{multline}
\begin{multline}
V_r(r)=-\mu^2 \left[Z_r(r)+Z_x(1)\right]-\frac{a^2 \left(m_{\phi }-2P_0 \,m_{\psi }\right)^2-I(r,0) \left[\omega-m_{\psi } \chi (r,0)\right]^2}{\Delta (r)}
\\-\frac{m_{\psi }^2\left[Z_r(r)+Z_x(1)\right]^3}{I(r,0)}-\left(m_{\phi }-2P_0\,m_{\psi}\right) \left[\omega\, \partial^2_\theta\omega _t(r,0)-m_{\psi }\,\partial^2_\theta\omega_{\psi}(r,0)\right]-\Lambda\,,
\end{multline}
\end{subequations}
where $\Lambda$ is the separation constant. As anticipated, the effective potentials $V_r(r)$ and $V_x(x)$ are closely related to the geodesic potentials $\Xi_r(r)$ and $\Xi_x(x)$. This correspondence reflects the \emph{eikonal limit}, in which the high-frequency behavior of wave propagation reduces to the study of null geodesics~\cite{Ferrari:1984zz}.
Specifically, upon identifying
\begin{subequations}
\begin{equation}
\sigma\leftrightarrow -\mu^2\,,\quad\mathcal{C}\leftrightarrow\Lambda\,,\quad m_{\phi}\leftrightarrow L_{\phi}\,,\quad m_{\psi}\leftrightarrow L_{\psi}\quad\text{and}\quad \omega\leftrightarrow E\,,
\end{equation}
then
\begin{equation}
V_r(r)=\Xi_r(r)\quad\text{and}\quad V_x(x)=\Xi_x(x)\,.
\end{equation}
\end{subequations}

\section{Closing comments}
\label{sec:discussion}

In this paper, we have constructed a new class of rotating generalizations of the static topological stars of~\cite{Bah:2020ogh,Bah:2020pdz}. Building such rotating solutions posed a significant challenge: they had to remain smooth in the interior while asymptoting to the standard Kaluza-Klein structure at infinity. Meeting these requirements demanded turning on many additional parameters compared to the static case, making the final solutions substantially more intricate. In particular, the construction involves a KKm charge parameter $n$ and a sequence of nontrivial sigma-model transformations that convert the Kerr-Taub-bolt dipole $a$ into genuine angular momentum. Regularity conditions quantize the solutions in terms of three integers, thereby generating towers of coherent states in supergravity that include BPS states, static non-BPS states, and an infinite family of non-BPS rotating configurations.

We demonstrated that these rotating geometries possess an ergoregion in five dimensions, though this disappears upon Kaluza-Klein reduction to four dimensions, indicating that the ergoregion is only visible to probes with nontrivial dependence along the compact fifth dimension. The solutions exhibit a sequence of spin-induced gravitational multipole moments that, while structurally analogous to those of the STU black hole, differ in important quantitative ways. Most strikingly, for certain topological stars, the multipoles can flip sign relative to those of black holes. For example, some topological stars display a positive spin-induced mass quadrupole, implying an exotic prolate shape elongated along the rotation axis, sharply contrasting with the familiar oblate profiles of ordinary rotating objects. Despite the intricate form of the solutions, we proved that they retain all the key features needed for a tractable dynamical analysis. In particular, both the geodesic equations and the Klein-Gordon equation separate into integrable radial and angular parts, ensuring that their gravitational signatures and stability properties can be derived analytically.

Finally, we established that rotating topological stars, unlike their static counterparts, do not coexist with black holes: their mass, charges, and angular momentum lie outside the black hole extremality bound of STU supergravity, even if the solutions satisfy the BPS bound. Nonetheless, we identified a parameter regime in which they approach a vacuum solution in five dimensions, with conserved charges nearly identical to those of a highly boosted Kerr black string, aside from a set of negligible extra charges. This makes them phenomenologically compelling prototypes of coherent microstates, where the black string horizon is replaced by a smooth rotating bubble supported by electromagnetic flux. Anticipating the analysis of their gravitational signatures, we compared their multipole moments with those of the boosted Kerr black string and found that, remarkably, their mass quadrupoles agree to within about $10\%$. \\

This work opens several promising directions, closely aligned with those of~\cite{Chakraborty:2025ger}:
\begin{itemize}
\setlength\itemsep{-0.5em}
    \item \underline{Constructing rotating topological solitons inside the black-hole regime:} The rotating solutions built here join the (long) list of nonextremal, rotating, smooth, horizonless geometries outside the black-hole bound~\cite{Jejjala:2005yu,Giusto:2007tt,Bossard:2014yta,Bossard:2014ola,Bena:2015drs,Bena:2016dbw,Bossard:2017vii}. However, unlike earlier examples, the way forward to solve this issue is clear: one must apply the sigma-model transformations to a seed solution that (1) lies inside the black-hole bound, (2) is smooth in the interior and asymptotically Kaluza-Klein, and (3) has a dipolar structure capable of generating spin after the transformation. While the Kerr-Taub-bolt solution satisfies (2) and (3), it fails (1). Our approach can therefore be repeated with a better candidate seed, and we already have promising candidates in mind~\cite{Bah:2022yji,Bah:2023ows,Heidmann:2023kry,Dulac:2024cso,Bonnor1966AnES,Becerril:1990ek}.
    \item \underline{Constructing rotating $\cW$-solitons:} In~\cite{Chakraborty:2025ger,Dima:2025tjz}, a new family of static topological solitons, dubbed $\cW$-solitons, was obtained by applying a different set of sigma-model transformations (compared to those used for topological stars) to the Taub-Bolt geometry. A natural next step is to build their rotating counterparts by applying analogous transformations to the Kerr-Taub bolt. While these solutions will still lie outside the black-hole bound, they could nevertheless provide interesting solutions and have a simpler analytic form.
    \item \underline{Constructing rotating topological solitons using perturbation theory:} The method we employed to construct rotating topological stars via sigma-model transformations may not yield \emph{generic} solutions. In particular, we do not seem to find solutions with arbitrarily small angular momentum near the static solitons constructed in \cite{Bah:2020ogh,Bah:2020pdz}. A natural way to explore the existence of such solutions is through perturbation theory, which is particularly tractable thanks to the spherical symmetry of the background geometry. Similar techniques could also shed light on whether $\cW$-solitons with arbitrarily small angular momentum can exist, opening a path toward a deeper understanding of these novel objects.
    \item \underline{Implications of a positive mass quadrupole:} Certain topological stars constructed here exhibit a positive spin-induced mass quadrupole (though not those close to the black-hole bound). This is highly exotic: rotation makes the object elongated along the axis rather than flattened at the equator. We know of no form of ordinary or exotic matter that responds to spin and gravity in this way.\footnote{The only exception we are aware of are certain gravastar models~\cite{Uchikata:2016qku}, which however require a thin-shell distribution of matter.}
    It would be valuable to determine the root of this effect, which we suspect to be a nontrivial interplay between the topological bubble, rotation, and dyonic charges. Interestingly, current gravitational-wave data may even favor positive mass quadrupoles~\cite{LIGOScientific:2021sio}, though the precision is still far from providing a strict bound.
    \item \underline{Stability and ergoregions:} The presence of an ergoregion typically implies the existence of superradiant modes~\cite{Brito:2015oca} which, in the absence of a horizon, can trigger ergoregion instabilities~\cite{Friedman:1978ygc,1978RSPSA.364..211C,Cardoso:2005gj,Chowdhury:2007jx,Cardoso:2007az,Moschidis:2016zjy}. Intuitively, modes trapped in the ergoregion can acquire negative energy and undergo repeated amplification through superradiance, leading to exponential growth. 
    
    In~\cite{Chowdhury:2007jx}, such instabilities were interpreted not as a flaw of coherent black-hole microstates, but rather as a manifestation of their strong coherence. Black holes are thermodynamically unstable and radiate via Hawking emission, so their microstates must also decay. In highly coherent cases, this decay can appear as a classical instability. As an analogy, consider a gas of atoms emitting radiation. In typical states, atoms radiate independently from different excited levels, and the emitted photons follow a Planckian distribution. By contrast, in a coherent state where all atoms occupy the same excited level and decay into a common lower state, the de-excited atoms form a Bose-Einstein condensate. Bose enhancement then increases the probability of subsequent de-excitations, leading to radiation that behaves as a classical, monochromatic wave with exponentially growing amplitude, i.e., a classical instability.

The question of the endpoint of such an instability is challenging, even for simpler black-hole geometries exhibiting Gregory-Laflamme instabilities~\cite{Dias:2009iu,Dias:2010eu,Figueras:2022zkg}. It becomes even more involved for the more complex geometries considered here. If only classical degrees of freedom are involved,
the system is likely to evolve toward a black-hole configuration \cite{Dias:2011ss,Niehoff:2015oga,Eperon:2016cdd,Eperon:2019viw,Siemonsen:2025ucx}. More realistically, however, and as in the case of a gas of excited atoms, the endpoint should be a more typical black-hole microstate reached through nontrivial transitions.

It would be interesting to perform an analysis similar to that of~\cite{Chowdhury:2007jx} to study the instability of rotating topological stars. Compared to JMaRT, the absence of a known CFT dual undermines the direct link to Hawking radiation made in~\cite{Chowdhury:2007jx}, but the fact that the ergoregion appears only for probes along the fifth dimension suggests that any instability will be highly suppressed, provided the extra dimension is small (e.g., of the order of a few Planck lengths).
    \item \underline{Gravitational signatures and black-hole comparison:} We have shown that both geodesic motion and scalar field dynamics are integrable and separable. This sets the stage for computing several phenomenological observables~\cite{Cardoso:2019rvt}, including photon scattering, photon rings, imaging simulations, scalar quasinormal modes, tidal Love numbers, and ringdown signals for rotating topological stars, particularly those close to the boosted Kerr black string, enabling direct phenomenological comparisons. While similar analyses have been performed for supersymmetric microstate geometries~\cite{Bena:2019azk,Bena:2020yii,Mayerson:2023wck,Bacchini:2021fig,Mayerson:2020tpn,Ikeda:2021uvc,Bianchi:2022qph} and more recently for static topological solitons~\cite{Lim:2021ejg,Bah:2021jno,Heidmann:2022ehn,Guo:2022nto,Guo:2022rms,Heidmann:2023ojf,Bianchi:2023sfs,Dima:2024cok,Bianchi:2024vmi,Dima:2025zot,Dima:2025tjz}, this would be the first time that a non-BPS coherent state from string theory is compared to an realistic nonextremal rotating black hole.
    \item \underline{Gravito-electromagnetic perturbations:} Ultimately, the full dynamics of rotating topological stars requires studying gravito-electromagnetic perturbations, which are significantly more challenging than scalar perturbations. The presence of a Killing tensor may indicate the existence of a Killing-Yano tensor for our rotating topological stars. The latter, if present, guarantees the separability of test Maxwell and Proca fields in such backgrounds~\cite{Frolov:2018ezx}. However, it remains unclear whether the same framework can be extended to probe spin-two fields, and even less so for gravito-electromagnetic perturbations. In such cases, one will likely need to employ a fully non-separable analysis, akin to the methods developed for studying the stability of Myers-Perry black holes~\cite{Dias:2009iu,Dias:2010maa,Dias:2011jg,Dias:2014eua}, black rings~\cite{Santos:2015iua}, and Kerr-Newman black holes~\cite{Dias:2015wqa,Dias:2021yju}, as well as for investigating perturbations driven by massive spin-two excitations on Kerr black holes~\cite{Dias:2023ynv}.
\end{itemize}

\section*{Acknowledgments}

We are deeply grateful to Iosif~Bena for his many insightful comments throughout the project and for the stimulating discussions that greatly enriched our work. We are also thankful to Massimo~Bianchi, Soumangsu~Chakraborty,  Jose~F.~Morales, Alejandro~Ruipérez for interesting discussion. We thank Marco Melis for pointing out several typos in equations. The work of P.\ H.\ is supported by the Department of Physics at The Ohio State University. P.\ P.\ is partially supported by the MUR FIS2 Advanced Grant ET-NOW (CUP:~B53C25001080001) and by the INFN TEONGRAV initiative. J.\ E.\ S.\ has been partially supported by STFC consolidated grant ST/X000664/1 and by Hughes Hall College. 

\appendix
\section{Sigma model of five-dimensional supergravity}
\label{sec:SigmaModel5d}

In this section, we summarize the sigma model emerging from five-dimensional STU supergravity and its solution-generating techniques as derived in~\cite{Chakraborty:2025ger}. Moreover, we restrict to a consistent truncation to minimal supergravity, determined by the action~\eqref{eq:L5min}.

\subsection{Sigma model}

We consider stationary solutions with two commuting $\text{U}(1)$ isometries, generated by $\partial_t$ (timelike) and $\partial_\psi$ (spacelike), for which the general ansatz takes the form
\begin{equation}
\begin{split}
    ds_5^2 &\= - \frac{1}{Z^{2}} \left( dt + \mu (d\psi + \omega_\psi) +\omega_t\right)^2 +Z \left[\frac{1}{Z_0} (d\psi + \omega_\psi)^2 + ds_3^2  \right]\,, \\
   A &\= A_t \, \left( dt + \mu (d\psi + \omega_\psi) +\omega_t\right)+A_\psi (d\psi + \omega_\psi) + \omega_B\,, 
   \label{eq:5dAnsatz}
\end{split}
\end{equation}
where the scalars $(Z_0, Z, \mu, A_t, A_\psi)$ and the one-forms $(\omega_t, \omega_\psi, \omega_B)$ are defined over the three-dimensional base space $ds_3^2$. Upon dualizing the one-forms into scalar potentials $(\Omega_t, \Omega_\psi, B)$, the five-dimensional theory reduces to a three-dimensional non-linear sigma model governed by the action~\cite{Chakraborty:2025ger}:
  \begin{equation}
  \label{eq:L3Matrix}
\cS_3 \=  \frac{1}{16 \pi G_3} \int \left( R_3 \star 1+ \frac{1}{8} \text{Tr} \left[d\cM^{-1} \wedge \star d\cM \right] \right)\,.
 \end{equation}
where $\cM$ is an $8\times 8$ coset matrix built from the eight scalar fields. The explicit form of $\cM$ in terms of the physical fields can be found in~\cite{Chakraborty:2025ger}. Additionally, one can define a matrix-valued one-form $\cN$ encoding the one-form fields:
\begin{equation}
d\cN \equi \cM^{-1} \star d\cM\,.
\label{eq:NMatrixDef}
\end{equation}

\subsection{Target-space transformations}

In STU supergravity, the target space of the sigma model is the coset $\text{SO}(4,4)/(\text{SO}(2,2)\times \text{SO}(2,2))$. This symmetry allows to generate new solutions by acting on a seed solution $(\cM, \cN, ds_3^2)$ with a group element $g \in \text{SO}(4,4)$:
\begin{equation}\label{eq:SOTransformation}
\cM \,\to\,  g^{T} \cM g\,,\qquad \cN \,\to\,  g^{-1} \cN g \,,\qquad ds_3^2 \,\to\,ds_3^2 \,.
\end{equation}
These transformations mix the scalar fields while leaving the base metric invariant, thus producing new supergravity solutions from a known seed.

For minimal supergravity, the coset symmetry reduces to $G_{2}$~\cite{Mizoguchi:1999fu,Clement:2007qy}. To retain compatibility with the techniques of~\cite{Chakraborty:2025ger}, we continue to work in the $\text{SO}(4,4)$ formalism and define the restricted subgroup for minimal supergravity, isomorphic to $G_{2}$.

\subsection{Reconstruction of five-dimensional fields}

The physical fields in the ansatz~\eqref{eq:5dAnsatz} can be explicitly extracted from the coset matrices $\cM$ and $\cN$ as follows:
\begin{align}
&Z_0 \=\sqrt{\cM_{42}^2 - \cM_{44} \cM_{22}}\,,\quad Z  \= -\frac{\cM_{44}}{\sqrt{\cM_{42}^2 - \cM_{44} \cM_{22}}},\quad \mu \= \frac{\cM_{32} \cM_{44} - \cM_{42} \cM_{43}}{\cM_{42}^2 - \cM_{44} \cM_{22}}, \nn\\
&A_t \= -\frac{\cM_{42}}{\cM_{44}}\,,\quad A_\psi \=\frac{\cM_{32} \cM_{42} - \cM_{22} \cM_{43}}{\cM_{42}^2 - \cM_{44} \cM_{22}}, \quad \omega_t \= -\cN_{74}\,,\quad \omega_\psi \= -\cN_{64}\,,\quad \omega_B \= -\cN_{54} \nn
\end{align}
It is also useful to define the combination that appears naturally upon dimensional reduction along the $\psi$ direction
\begin{equation}
\cI_4 \equi Z_0 Z^3  - \mu^2 Z_0^2 \=\cM_{44}\cM_{33}-\cM_{43}^2.
\label{eq:QuarticDef}
\end{equation}

\subsection{$G_2$ transformations preserving asymptotic flatness}
\label{sec:SO(4,4)Transf}

We focus on constructing solutions that asymptote to a Kaluza-Klein background:\footnote{The addition of potential NUT or magnetic charges does not affect the discussion.}
\begin{equation}
ds_{5,\text{asym}}^2 \= -dt^2 + d\psi^2 + dr^2 + r^2 (d\theta^2 +\sin^2 \theta\,d\phi^2)\,,
\label{eq:AsympR13S1}
\end{equation}
As established in~\cite{Chakraborty:2025ger}, the subgroup of $\text{SO}(4,4)$ preserving this asymptotic structure is twelve-dimensional. Its intersection with $G_{2}$ yields a six-dimensional subgroup, which decomposes into three commuting parts: the $\cP$-, $\cZ$-, and $\cW$-groups. We refer the reader to Section 3.1 of~\cite{Chakraborty:2025ger} for a detailed definition of the $\mathfrak{so}(4,4)$ generators.
\begin{itemize}
\item[-] \underline{$\cP$-group:} This group is generated by two Lie algebra elements, $\cP = \sum_{I=1}^3 (\cP_{+I}+\cP_{-I})$ and $\cX = \cX_+ + \cX_-$.  A generic element takes the form
\begin{equation}
g_\cP \= \exp \left[ \delta_1 \,\cP + \delta_2 \,\cX \right],
\label{eq:Pgroup}
\end{equation}
where $(\delta_1,\delta_2)$ are two arbitrary constants.
\item[-] \underline{$\cZ$-group:} Generated by $\cZ = \sum_{I=1}^3 (\cZ_{+I}+\cZ_{-I})$ and $\cO_1 = \cO_{+1}+ \cO_{-1}$, its elements take the form
\begin{equation}
g_\cZ \= \exp \left[ \gamma_1 \,\cZ + \gamma_2 \,\cO_{1}\right],
\label{eq:Zgroup}
\end{equation}
with parameters $(\gamma_1,\gamma_2)$.
\item[-] \underline{$\cW$-group:} This group is generated by $\cW = \sum_{I=1}^3 (\cW_{+I} - \cW_{-I})$ and $\cO_2 = \cO_{+2} - \cO_{-2}$. A typical group element is
\begin{equation}
g_\cW \= \exp \left[  \alpha_1 \,\cW + \alpha_2 \,\cO_{2}\right],
\label{eq:Wgroup}
\end{equation}
with parameters $(\alpha_1,\alpha_2)$.
\end{itemize}

\subsection{Solution-generating technique}
\label{sec:SolGenTec}

By acting with the $\cP$, $\cZ$, and $\cW$ transformations on a suitable seed, one can generate a rich class of smooth, non-extremal, rotating geometries. The procedure proceeds as follows:
\begin{itemize}
    \item Begin with a smooth, horizonless seed solution with asymptotics $\IR^{1,3} \times \text{S}^1$.
    \item Apply a composition of $\cP$, $\cZ$, and $\cW$ transformations.
    \item Impose the absence of a NUT charge along the time direction for the transformed solution:
    \begin{equation}
        \omega_t \underset{r\to\infty}{\to} 0\,.
        \label{eq:NoNUT}
    \end{equation}
    \item Since the base metric $ds_3^2$ is invariant under the transformations, all coordinate degeneracies in the seed persist in the transformed solution. One must tune the transformation parameters to ensure that these degeneracies correspond to smooth loci where an angular direction shrinks smoothly.

More precisely, consider a generic base, $ds_3^2=g_{rr} dr^2+g_{\theta \theta} d\theta^2+g_{\phi\phi} d\phi^2$, with one-forms along $\phi$, namely $(\omega_t, \omega_\psi) = (\omega_t, \omega_\psi)\, d\phi$. If $r=r_0$ is a spacelike coordinate degeneracy in the seed solution, it remains such a locus in the transformed solution provided that the determinant of the induced metric along $(\psi,\theta,\phi)$ vanishes at $r=r_0$:
    \begin{equation}
        \det g_{(\psi \theta \phi)}|_{r=r_0} \= g_{\theta \theta} \left( g_{\phi \phi} Z_0^2 \cI_4 - \omega_t^2\right) |_{r=r_0} =0\,.
        \label{eq:RegBolt}
    \end{equation}
    \item Finally, physical viability of the solution requires a positive ADM mass and the absence of CTCs. The latter is guaranteed by ensuring that $g^{tt} < 0$ everywhere, so that $t$ is a global time function.
\end{itemize}

\section{The Kerr-Taub-bolt as a seed solution} \label{sec:Kerr-Taub-bolt}

In this section, we analyze in detail the seed geometry that is used to construct rotating topological stars via $G_{2}$ transformations: the five-dimensional uplift of the Kerr-Taub-bolt solution. While the transformations act nontrivially on the fields of the theory, they partially preserve the spacetime structure. A detailed examination of the spacetime properties and regularity conditions of the seed solution is therefore valuable, and it is the focus of this section.

\subsection{The Kerr-Taub-bolt}

The Kerr-Taub bolt is a vacuum solution of five-dimensional gravity. It is obtained by Wick rotating the Kerr-NUT spacetime and trivially adding a time direction. To preserve reality after the Wick rotation, the spin and NUT parameters must also be analytically continued: $(a, n) \rightarrow i(a, n)$. The resulting solution is given in~\eqref{eq:KerrTaubBolt}. The parameters $(m, n, a)$ directly encode the asymptotic charges: $m$ sets the ADM mass, $n$ the Kaluza-Klein monopole charge, and $a$ the KK dipole moment. However, for analyzing regularity and infrared structure, it is more convenient to parameterize the solution in terms of $(r_+, r_-, a)$~\eqref{eq:m&nrpm}.

Several important limits are worth highlighting:
\begin{itemize}
    \item \underline{Extremal limit:} The base~\eqref{eq:KerrTaubBolt} becomes Ricci-flat when
    \begin{equation}
        2|a| \= r_+-r_-\,. \label{eq:ExtremalLimitApp}
    \end{equation}
    the geometry reduces to the BPS Taub-NUT solution. For instance, choosing $2a = r_- - r_+$ and $r_+ + r_- > 0$,\footnote{We could have considered other allowed values, which will have simply changed which of the North or South pole coordinates we need to make the Taub-NUT structure manifest.} and introducing coordinates centered around the North pole
    \begin{equation}
\begin{split}
r &\= \frac{1}{2}\left[r_{N} +r_++r_- + \sqrt{(r_{N}\cos \theta_{N}+r_+-r_-)^2+r_{N}^2 \sin^2 \theta_{N}} \right] \\
\cos \theta &\=   \frac{\sqrt{(r_{N}\cos \theta_{N}+r_+-r_-)^2+r_{N}^2 \sin^2 \theta_{N}} -r_{N}}{r_+-r_-}\,,
\end{split}
\end{equation}
the solution becomes
\begin{equation}
    ds^2 \= -dt^2 +\frac{1}{Z_0} \left( d\psi+(r_++r_-) (\cos \theta_N+1)\right)^2+Z_0 \left[dr_N^2 +r_N^2 \left(d\theta_N^2 + \sin^2 \theta_N \,d\phi^2 \right)\right],  \nn
\end{equation}
where $Z_0 = 1 + \frac{r_+ + r_-}{r_N}$.
    \item  \underline{``Static'' limit ($a = 0$):} The solution becomes spherically symmetric, and the KK dipole vanishes. Although the original geometry is already static for arbitrary $a$, this limit corresponds to the static limit of the Kerr-NUT black hole before Wick rotation. This yields the Taub-bolt geometry plus a time direction:
    \begin{equation}
        ds_{a=0}^2 \= -dt^2 + \frac{\Delta}{r^2-r_+ r_-} \left[d\psi+2\sqrt{r_+ r_-} (\cos\theta+1) d\phi \right]^2 +\frac{r^2-r_+ r_-}{\Delta} \,ds_3^2|_{a=0},\label{eq:TaubBolt}
    \end{equation}
    where $\Delta=(r-r_+)(r-r_-)$ and $ds_3^2|_{a=0}=dr^2+\Delta (d\theta^2+\sin^2 \theta d\phi^2)$.
\end{itemize}

\subsection{Key condition on the asymptotic structure}
\label{sec:CondAsympSmooth}

At large distances, $r\gg r_\pm$, the solution asymptotes to 
\begin{equation}
    ds^2 \,\to\, -dt^2 + [d\psi+ 2n(\cos \theta +1) d\phi]^2 + dr^2 + r^2 \left(d\theta^2+\sin^2 \theta\,d\phi^2\right)\,. \label{eq:AsymGen}
\end{equation}
 Contrary to the claim made in~\cite{Bianchi:2025uis}, this asymptotic form does not necessarily describe the standard Kaluza-Klein geometry, i.e., a trivial S$^1$ fibration over four-dimensional Minkowski space. The global structure at infinity crucially depends on the periodic identifications imposed on the angular coordinates $(\psi,\phi)$.

If we assume the standard periodicity lattice,
\begin{equation}
(\phi,\psi) \= (\phi,\psi) + (2\pi,0)\,,\qquad (\phi,\psi) \= (\phi,\psi) + (0,2\pi R_\psi)\,,
\label{eq:AngleIdentificationApp}
\end{equation}
with $R_\psi$ the radius of the S$^1$, then $\partial_\psi$ is a globally defined Killing vector with closed orbits. In this frame, $\phi$ remains constant along $\partial_\psi$, and one can consistently perform a Kaluza-Klein reduction along $\psi$ at fixed $\phi$. The resulting geometry is asymptotically a compact S$^1$ fibered over flat $\mathbb{R}^{1,3}$.

However, if, as in~\cite{Bianchi:2025uis,Dowker:1995gb}, the periodicity lattice is
\begin{equation}
(\phi,\psi) \= (\phi,\psi) + (2\pi,0)\,,\qquad (\phi,\psi) \= (\phi,\psi) + (2\pi \alpha,2\pi R_\psi)\,,\qquad |\alpha| <1\,,
\label{eq:AngleIdentificationBad}
\end{equation}
then the Killing vector with closed orbits becomes $\partial_\psi + \bar{\alpha} \partial_\phi$, where $\bar{\alpha} \equiv \alpha/R_\psi$. One must introduce a new coordinate $\widetilde{\phi} \equiv \phi - \bar{\alpha} \psi$, which remains constant along these orbits and satisfies the standard identifications~\eqref{eq:AngleIdentificationApp}. Rewriting the metric in terms of $\widetilde{\phi}$, we obtain:
\begin{equation}
    ds^2 \,\to\, -dt^2 + \Lambda (d\psi+ \omega_\psi^\infty \, d\widetilde{\phi})^2 + dr^2 + r^2 \left(d\theta^2+\frac{\sin^2 \theta}{\Lambda}\,d\widetilde{\phi}^2\right)\,,
\end{equation}
where $$\Lambda \to \bar{\alpha}^2 r^2 \sin^2 \theta 
,\qquad \omega_\psi^\infty  \to \bar{\alpha}^{-1}
. $$

The key observation is that $\Lambda$ grows with $r$, indicating that the $\psi$ circle decompactifies at infinity. Consequently, the space does not asymptote to a Kaluza-Klein geometry. Upon dimensional reduction, the resulting four-dimensional geometry is:
\begin{equation}
    ds_4^2 \= \sqrt{\Lambda} \left(-dt^2+dr^2+r^2 d\theta^2 \right) + \frac{r^2 \sin^2 \theta}{\sqrt{\Lambda}} d\widetilde{\phi}^2\,,\quad e^{-\frac{4}{\sqrt{3}} \Phi} \= \Lambda\,, \quad A_{KK} = \omega_\psi^\infty\,d\widetilde{\phi}\,.
\end{equation}
Asymptotically, this behaves as $ds_4^2 \to \bar{\alpha}\,r\, \sin \theta \left[-dt^2 +dr^2+r^2d\theta^2+ \frac{d\widetilde{\phi}^2}{\bar{\alpha}^2} \right],$ corresponding to a non-Minkowski geometry with a vanishing scalar field ($\Phi \to 0$) and constant magnetic flux $A_{KK} \to \frac{1}{\bar{\alpha}}\, d\widetilde{\phi}$. This class of solutions, known as \emph{magnetic flux tubes} or \emph{Melvin universes}~\cite{Melvin:1963qx}, was studied in detail in~\cite{Dowker:1995gb}. \\

In summary, for the solution to be asymptotic to a compact S$^1$ over Minkowski space, two conditions must be satisfied: the metric must approach~\eqref{eq:AsymGen}, \emph{and} the angular periodicities must be those in~\eqref{eq:AngleIdentificationApp}. The periodicities of the form~\eqref{eq:AngleIdentificationBad} are incompatible with Kaluza-Klein asymptotics, rendering the solution unrealistic from a phenomenological point of view.

\subsection{Conserved charges and black-hole regime}

Assuming the proper periodicity lattice~\eqref{eq:AngleIdentificationApp}, the solutions asymptote to $\mathbb{R}^{1,3} \times$S$^1$, and a consistent four-dimensional Kaluza-Klein reduction along $\psi$ can be performed. The resulting fields are:
\begin{equation}
\begin{split}
    ds_4^2 = - \sqrt{\frac{\Delta +a^2 \sin^2 \theta}{\Sigma}} dt^2 + \sqrt{\frac{\Sigma}{\Delta +a^2 \sin^2 \theta}}\,ds_3^2,\quad A_{KK} = \omega_\psi \,d\phi,\quad e^{-\frac{4}{\sqrt{3}} \Phi} = \frac{\Delta +a^2 \sin^2 \theta}{\Sigma}.
\end{split}
\end{equation}
This corresponds to a static, massive magnetic configuration whose ADM mass $M$ and Kaluza-Klein monopole (KKm) charge $P_0$ (in $G_4=1$ units) are:
\begin{equation}
    P_0 \= 2n\,,\qquad M\= \frac{m}{2}\,.
\end{equation}
The dipole parameter $a$ is related to the magnetic dipole moment: $\mathcal{J}_m = am$.

A necessary condition for the existence of a regular solution is that $r_\pm$ are real, i.e., $m^2 + a^2 - n^2 \geq 0$. In terms of charges and the magnetic dipole, this becomes:
\begin{equation}
    M^2 + \frac{\cJ_m^2}{16M^2}-\frac{P_0^2}{16} \,\geq \,0\,.
    \label{eq:ValidityBound}
\end{equation}
Because $\mathcal{J}_m$ is not a conserved quantity, solutions exist for any values of $(M, P_0)$ provided the magnetic dipole is suitably tuned.

Moreover, we will see in the next section that $\Sigma$ becomes negative in certain regions when $|n| > m$ or, equivalently, $2|a| < r_+ - r_-$. In this parameter range, the Kerr-Taub-bolt geometry becomes ambipolar, undergoing a change in signature from $(-,+,+,+,+)$ to $(-,-,-,-,-)$. As a result, the four-dimensional reduction becomes ill-defined. Therefore, a well-defined Kerr-Taub-bolt geometry requires the additional condition:
\begin{equation}
    M \geq  \frac{|P_0|}{4}.
\end{equation}

In the extremal limit, $M = \frac{|P_0|}{4}$, the solution reduces to the BPS Taub-NUT geometry, in agreement with the BPS bound for KK monopoles in five dimensions.

Finally, the Kerr-Taub-bolt has the same conserved charges as the static Kaluza-Klein magnetic black hole:
\begin{align}
    ds^2 &\= - \Delta\,dt^2+\frac{1}{Z_0} \left(d\psi+P_0(1+\cos \theta)d\phi \right)^2 + Z_0 \left[\frac{dr^2}{\Delta}+r^2 \left(d\theta^2+\sin^2 \theta\,d\phi^2 \right)\right],\nn\\
    \Delta &\= 1- \frac{3M-\sqrt{M^2+\frac{P_0^2}{2}}}{r},\qquad Z_0 \= 1+\frac{2\left(\sqrt{M^2+\frac{P_0^2}{2}}-M\right)}{r}.
\end{align}
The black hole exists whenever $M > |P_0|/4$, and the solution becomes the BPS Taub-NUT geometry when $M = |P_0|/4$. Thus, Kerr-Taub-bolt solutions coexist with black holes of identical conserved charges when $M > |P_0|/4$.

\subsection{Regularity in the interior}
\label{sec:RegKerrTaubBolt}

The Kerr-Taub-bolt geometry has five special loci. Four correspond to coordinate degeneracies: two at the poles of the two-sphere ($\theta = 0, \pi$) and two at the radial surfaces $r = r_\pm$ where $\Delta=0$. One corresponds to a singularity where the four-dimensional base changes signature from $(+,+,+,+)$ to $(-,-,-,-)$ when $\Sigma\leq 0$.

\begin{itemize}
    \item \underline{Ambi-polar condition:}
\end{itemize}

The base changes signature if the largest zero of $\Sigma$ occurs before the coordinate degeneracy at $\Delta = 0$. Since $\Sigma = 0$ when $r = \pm(n + a \cos \theta)$, the base becomes ambipolar if $|n| + |a| > r_+$. This leads to the following conditions:
\begin{equation}
    \begin{split}
        &|n| \,>\, m \quad (2|a| > r_+ - r_-)\quad \Leftrightarrow \quad \text{The base is ambipolar with regions where $\Sigma \leq 0$.}\\
        &|n| \,\leq\, m \quad (2|a| \leq r_+ - r_-)\quad \Leftrightarrow \quad \text{The base does not change signature with $\Sigma > 0$.}
    \end{split}
    \label{eq:AmbipolarCond}
\end{equation}

\begin{itemize}
    \item \underline{At the poles of the two-sphere:}
\end{itemize}

At $\theta = 0$ and $\theta = \pi$, the $\phi$ circle degenerates at fixed $\psi + \omega_\psi |_{\theta=0,\pi} \phi$. Near these poles, the metric reduces to $d\theta^2 + \sin^2\theta\, d\phi^2$, which is smooth if $\phi$ has periodicity $2\pi$. We have already chosen a gauge in which $\omega_\psi|_{\theta=\pi} = 0$, so the standard identification~\eqref{eq:AngleIdentificationApp} ensures regularity at $\theta = \pi$. At $\theta=0$, the $\phi$ direction shrinks at fixed $\psi + 4n \phi = \psi + 2P_0 \phi$. The identification is $2\pi$ only if $2P_0 \in \mathbb{Z}\, R_{\psi}$. Thus, the identification is smooth if
\begin{equation}
P_0 \= \frac{1}{2} N R_\psi\,,\qquad N \in \mathbb{Z}\,.
\end{equation}

\begin{itemize}
\item[•] \underline{At the bolt:}
\end{itemize}

Near $r = r_+$, we introduce the local radial coordinate $\rho$~\eqref{eq:CoorBolt},
and expand the $(\rho,\phi,\psi)$ part of metric near $\rho = 0$, yielding the same metric as for the rotating topological star~\eqref{eq:MetricBolt}, with:
\begin{equation}
\omega_\psi |_{r=r_+} \= -2 a +\frac{r_+(r_+-r_-)}{a}\,.
\end{equation}
We similarly introduce local angles $\widetilde{\phi}$ and $\widetilde{\psi}$,~\eqref{eq:LocalAngles}, for which the global identification~\eqref{eq:AngleIdentificationApp} leads to
\begin{equation}
\left(\widetilde{\phi},\widetilde{\psi} \right) \= \left(\widetilde{\phi},\widetilde{\psi} \right) + \begin{cases} 
2\pi \left( \frac{r_+-r_-}{2|a|} ,  \omega_\psi|_{r=r_+} \right)  \qquad \qquad &(A)\\
2\pi \left(0 , R_\psi \right) \qquad &(B)
\end{cases},
\end{equation}
As argued in Section~\ref{sec:RegCondTS}, the shrinking direction at the bolt is $\widetilde{\phi}$ at fixed $\widetilde{\psi}$, such that the bolt is smooth provided the following conditions hold:
\begin{align}
&(1)\quad \omega_\psi|_{r=r_+} \= \frac{\ell}{k} \, R_\psi\,, \qquad (\ell,k) \in \mathbb{Z}\,,\qquad \gcd(\ell,k)=1,\qquad k\geq 0, \label{eq:RegCondRpsi2} \\
&(2) \quad |a| \= \frac{k\,(r_+ - r_-)}{2}\,. \label{eq:RegCondApp}
\end{align}
The second condition quantizes the spin parameter $a$ into a tower of allowed values labeled by an integer $k$, which also translates into a condition on the KKm charge:
\begin{equation}
    n \= \pm \sqrt{m^2+a^2\left(1-\frac{1}{k^2}\right)}\,.
    \label{eq:nFixKTBApp}
\end{equation}
Finally, using~\eqref{eq:RegCondRpsi2} and the quantization of the KKm charge, we can express $R_\psi$ and $\ell$ in terms of the seed parameters and $k$:
\begin{equation}
    R_\psi \= \frac{4\sqrt{a^2+r_+r_-}}{N},\qquad \frac{2\ell}{k N} \= \frac{r_+^2-\left(a-\sqrt{a^2+r_+r_-}\right)^2}{2 a\sqrt{a^2+r_+r_-}}.
    \label{eq:RegKerrTaubBolt}
\end{equation}
This regularity condition~\eqref{eq:nFixKTBApp} has several important consequences:
\begin{itemize}
    \item If $a \neq 0$, no regular solutions exist with $n = 0$. Therefore, the Kerr-bolt geometry (i.e., the Kerr-Taub-bolt with $n = 0$) used in~\cite{Bianchi:2025uis} is not a suitable seed geometry to construct solutions that are at the same time smooth at the bolt and asymptotic to $\text{S}^1 \times \IR^{1,3}$.
    \item For $k \geq 2$, regularity at the bolt requires $|n| > m$, and the Kerr-Taub-bolt becomes ambipolar~\eqref{eq:AmbipolarCond}. Consequently, a Kerr-Taub-bolt geometry with the correct periodicity lattice~\eqref{eq:AngleIdentificationApp} cannot simultaneously be regular at the bolt and free of signature change in the interior. Moreover, such geometries require magnetic charges exceeding the BPS bound, 
    \begin{equation}
        M \,\leq\, |P_0|/4,
        \label{eq:OutsideBHboundKTB}
    \end{equation}
    meaning that no corresponding black hole solutions exist with these values of mass and charges.
    \item The $k = 1$ geometry corresponds to the extremal BPS solution~\eqref{eq:ExtremalLimitApp} with $|P_0| = 4M$.
    \item The $k = 0$ solution is well defined since $2|a| = k(r_+ - r_-) = 0$, and the Kerr-Taub-bolt reduces to the regular Taub-bolt geometry~\eqref{eq:TaubBolt}. One can verify that the regularity conditions~\eqref{eq:RegKerrTaubBolt} are well defined in this limit, leading to the Taub-bolt regularity conditions: $\ell R_\psi = 2 r_+$ and $\frac{4\ell^2}{N^2} = \frac{r_+}{r_-}$, with $|n| \leq m$, so that the solution is not ambipolar.
\end{itemize}
The analysis shows that the following three conditions cannot be satisfied simultaneously for solutions with $a\neq 0$: (1) the solution is asymptotic to $\text{S}^1 \times \IR^{1,3}$; (2) the solution is smooth at the bolt; and (3) the solution is free of signature change. However, since the Kerr-Taub-bolt geometry only serves as a seed for sigma-model transformations, one can still use it as a valid seed that satisfies conditions (1) and (2), conditions that cannot be modified by transformations, and then fine-tune the transformations that (3) is satisfied for the transformed geometry.

\section{Constructing rotating solutions from the Kerr-Taub-bolt}
\label{sec:ConstructionSteps}

As shown in the previous Appendix, the Kerr-Taub-bolt geometry can be made smooth at the cap and asymptotically Kaluza-Klein space when either $(n, a) \neq 0$ or $a = 0$. It is therefore a suitable seed for generating novel smooth, horizonless solutions in five-dimensional supergravity using $G_{2}$ transformations. These transformations are reviewed in Section~\ref{sec:SO(4,4)Transf}, and the solution-generating procedure is summarized in Section~\ref{sec:SolGenTec}.

To implement this, we first compute the coset matrices $(\cM_0, \cN_0)$ corresponding to the Kerr-Taub-bolt seed:
\begin{align}
    \cM_0 \= &\delta_{11} -\delta_{33}+\delta_{55}-\delta_{77}+ \frac{\Sigma}{\Delta+a^2\sin^2\theta} \left( \delta_{22}-\delta_{44}-\delta_{66}+\delta_{88} \right) \nn\\
    &- \frac{2R}{\Delta +a^2\sin^2 \theta} \left( \delta_{28}+\delta_{46}+\delta_{64}+\delta_{82}\right)+ \left(2+\frac{m^{-}_{1,0}{}^{2}-4a^2}{\Delta+a^2 \sin^2\theta} \right) \left(\delta_{66}-\delta_{88}\right),\nn\\
    \cN_0 \= & - \omega_\psi d\phi \left(\delta_{28}- \delta_{46}+\delta_{64}-\delta_{82} \right)  + \frac{(m^{-}_{1,0}{}^{2}-4a^2)a\sin^2\theta}{\Delta+a^2 \sin^2\theta} \left(\delta_{66}-\delta_{88}\right) \\
    &+ \left[(r_++r_- )\cos \theta+ \frac{ 2a R\, \sin^2\theta}{\Delta+a^2 \sin^2\theta} \right] d\phi \left(\delta_{22}+\delta_{44}-\delta_{66}-\delta_{88} \right) \,,\nn
\end{align}
where $\delta_{ij}$ denotes the $8\times8$ matrix with a 1 at entry $(i,j)$ and zeros elsewhere.

The structure of $\cN_0$ shows that the application of $G_{2}$ transformations maps the gravitational and magnetic monopole charges $(m,n)$, defined in~\eqref{eq:m&nrpm}, and the dipole parameter $a$ into new charges in five-dimensional supergravity. In particular, the magnetic dipole $a$ is responsible for introducing angular momentum, and thus for generating rotating solutions.

By applying the roadmap from Section~\ref{sec:SolGenTec}, one can systematically obtain solutions that are smooth and horizonless. Crucially, regularity must be preserved at the coordinate degeneracies inherited from the seed, which imposes specific constraints on the transformation parameters. These include the absence of NUT charge at infinity and the absence of pathologies at the bolt:
\begin{equation}
    \omega_t \underset{r\to \infty}{\to} 0\,,\qquad \begin{cases}
        \omega_t \underset{r\to r_+}{\to} 0\,,\qquad &\text{ if }a\neq 0\,,\\
        \mu \underset{r\to r_+}{\to} 0\,,\qquad &\text{ if }a\= 0\,.
    \end{cases}
    \label{eq:RegCondGenRot}
\end{equation}
Previously known solutions can be naturally obtained from these transformations:
\begin{itemize}
    \item \underline{Static topological star of~\cite{Bah:2020ogh}:} Obtained by applying a $\mathcal{P}$-group transformation~\eqref{eq:Pgroup} with $\delta_2 = 0$ to a Euclidean Schwarzschild seed, which corresponds to the $a = n = 0$ limit of the Kerr-Taub-bolt.
    \item  \underline{``Rotating topological star'' of~\cite{Bianchi:2025uis}:} Constructed by applying a $\mathcal{P}$-group transformation to a Kerr-bolt seed (the $n = 0$ limit of the Kerr-Taub-bolt).
\end{itemize}
However, as previously argued, the solution in~\cite{Bianchi:2025uis} is ultimately unsatisfactory: smoothness at the bolt is incompatible with Kaluza-Klein asymptotics when $n = 0$. To overcome this, we seek a smooth and asymptotically flat solution by allowing for nonzero $n$.

Surprisingly, we find that no smooth solution satisfying~\eqref{eq:RegCondGenRot} arises from a pure $\mathcal{P}$-group transformation when $n \neq 0$. To resolve this, we introduce an additional transformation. In this paper, we choose a $\mathcal{W}$-group transformation~\eqref{eq:Wgroup} with $\alpha_1 = 0$. This is not a unique choice as other $\mathcal{W}$ or $\mathcal{Z}$ transformations could also be used to produce alternative geometries. However, the transformation we implement provides the first explicit construction of a rotating topological star with $n \neq 0$ that is both smooth and asymptotically S$^1\times\IR^{1,3}$.\\

The asymptotically flat rotating topological star is obtained by applying the $\cP$ and $\cW$ transformation, $g = \exp [ \alpha_2 \cO_2] .  \exp [ \delta_1 \cP +\delta_2 \cX]$ to the Kerr-Taub-bolt solution. 

Compared to the static topological star of~\cite{Bah:2020ogh}, the new solution involves four additional parameters. Relative to the non-asymptotically flat rotating version of~\cite{Bianchi:2025uis}, it introduces two more. As a result, the final solution has a significantly more intricate structure than these earlier examples.

Imposing the regularity conditions~\eqref{eq:RegCondGenRot} fixes two transformation parameters:
\begin{equation}
    \tanh \delta_2 \= - \frac{r_-}{r_+} \tanh^3 \delta_1\,,\qquad \sin \alpha_2 \= \frac{2\sqrt{a^2+r_- r_+}\,\tanh^3 \delta_1}{r_++r_- \tanh^6 \delta_1}\,.
\end{equation}
The solution is thus characterized by four independent parameters $(r_+, r_-, a, q)$, where $|q| < 1$ and $q = \tanh \delta_1$.

\bibliographystyle{utphys}      

\bibliography{microstates}       

\end{adjustwidth}

\end{document}